\documentclass[11pt]{article}

\usepackage[final]{acl}

\usepackage{times}
\usepackage{latexsym}

\usepackage[T1]{fontenc}

\usepackage[utf8]{inputenc}

\usepackage{microtype}

\usepackage{inconsolata}

\usepackage{graphicx}

\usepackage{algorithm}
\usepackage{algorithmic}
\usepackage{makecell}
\usepackage{newfloat}
\usepackage{listings}
\usepackage{times}
\usepackage{epsfig}
\usepackage{float}
\usepackage{placeins}
\usepackage{enumitem}
\usepackage{tabularx}
\usepackage{xstring}
\usepackage{xspace}
\usepackage[hang,flushmargin]{footmisc}
\usepackage[utf8]{inputenc} 
\usepackage[T1]{fontenc}    
\usepackage{url}            
\usepackage{booktabs}       
\usepackage{amsfonts}       
\usepackage{nicefrac}       
\usepackage{xcolor}         
\usepackage{amsmath, amssymb, overpic, textpos}
\usepackage{graphicx}
\usepackage{multirow}
\usepackage{color, colortbl}
\usepackage{tabulary}
\usepackage{listings}
\usepackage{multicol}
\usepackage{xspace}
\usepackage{graphbox}
\usepackage{arydshln}
\usepackage{adjustbox}
\usepackage{kotex}
\usepackage{caption}
\usepackage{subcaption}
\usepackage{makecell}
\usepackage{tabularray}
\usepackage{bm}

\definecolor{Green}{rgb}{0.2, 0.7, 0.1}
\definecolor{Gray}{HTML}{B0B0B0} 
\definecolor{Blue}{HTML}{4A90E2} 
\definecolor{Yellow}{HTML}{FFA700} 
\definecolor{prompt_blue}{HTML}{1f78b4}
\definecolor{prompt_red}{HTML}{d45c43}
\definecolor{plus}{HTML}{0071bc}
\definecolor{minus}{RGB}{153,10,10}
\definecolor{SecondBest}{HTML}{E0F0FA}
\definecolor{Best}{HTML}{BAD8F2}

\usepackage[table]{xcolor}
\usepackage[utf8]{inputenc}

\usepackage{tcolorbox}
\usepackage{etoc}
\usepackage{titletoc}

\title{Generalizable Prompt Tuning for Audio-Language Models \\ via Semantic Expansion}

\author{
    Jaehyuk Jang\thanks{Equal contribution} \qquad
    Wonjun Lee\footnotemark[1] \qquad
    Kangwook Ko\footnotemark[1] \qquad
    Changick Kim\\
    \textsuperscript{} KAIST \\
    {\tt\small \{jhyuk, dpenguin, kw.ko, changick\}@kaist.ac.kr}\\
}

\begin{document}
\maketitle

\begin{abstract}

Prompt tuning has achieved remarkable progress in vision–language models (VLMs) and is recently being adopted for audio–language models (ALMs).
However, its generalization ability in ALMs remains largely underexplored.
We observe that conventional prompt tuning for ALMs also suffers from the Base–New Tradeoff, and we identify that this issue stems from the disrupted semantic structure of the embedding space.
To address this issue, we propose \textbf{S}emantically \textbf{E}xpanded \textbf{P}rompt \textbf{T}uning (\textbf{SEPT})—a plug-and-play framework that explicitly regularizes the prompt embedding space by incorporating semantic neighbors generated by large language models.
SEPT introduces a novel semantic expansion loss with margin constraints that promote intra-class compactness and inter-class separability, thereby enhancing the semantic structure of the prompt embedding space.
For comprehensive evaluation, we establish the first benchmark setup for prompt generalization in ALMs, covering both base-to-new generalization and cross-dataset transferability.
Extensive experiments demonstrate that SEPT consistently improves generalization performance across multiple prompt tuning baselines, while maintaining computational cost during inference.

\end{abstract}

\label{sec:intro}
\section{Introduction}

\begin{figure}[t]
    \centering
    \includegraphics[width=\linewidth]{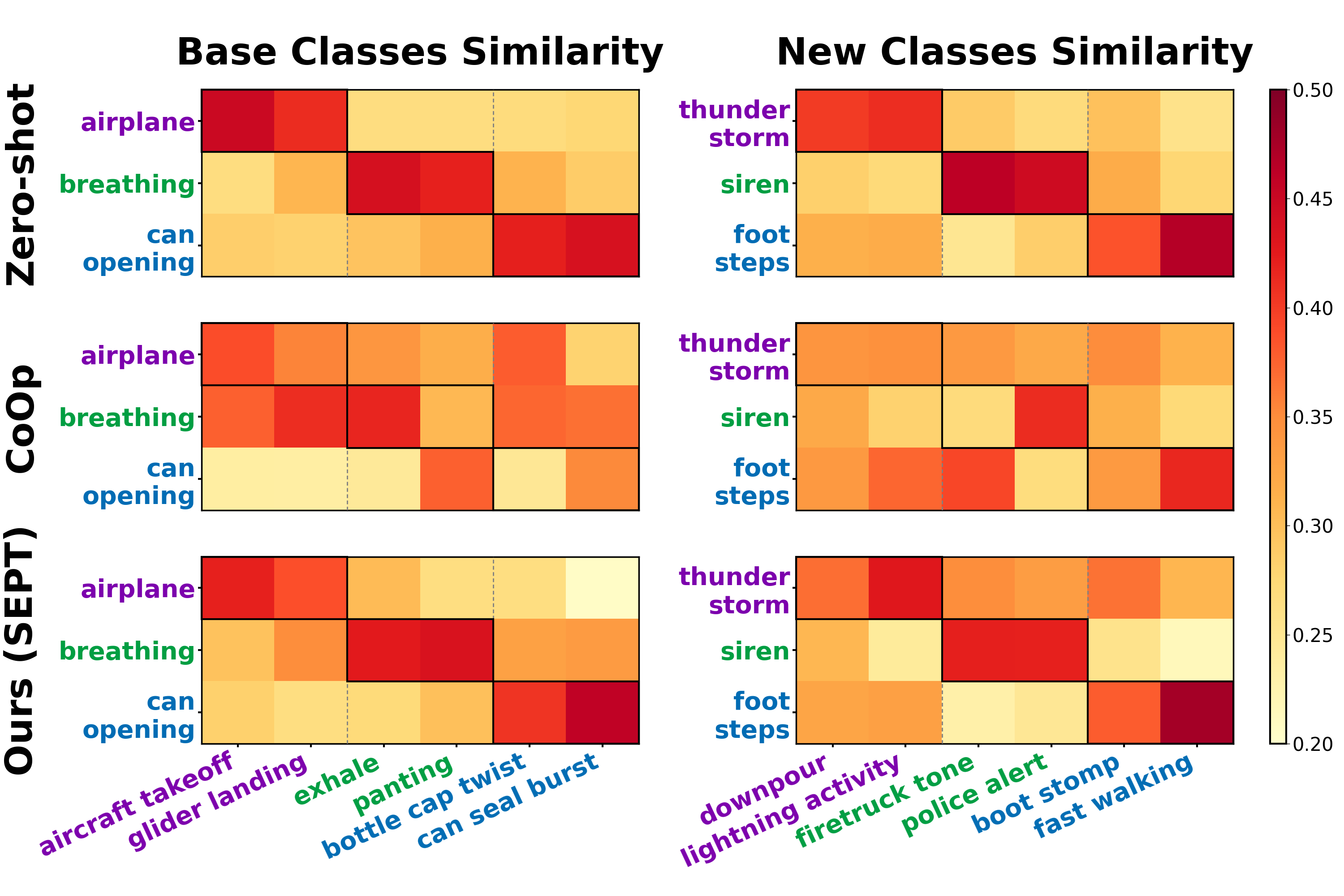}
    \caption{
    Motivation. The $y$-axis corresponds to class names, and $x$-axis lists semantic neighbors for each class. Hand-crafted zero-shot prompts show strong similarity between each class and its neighbors. In contrast, conventional prompt tuning disrupts this alignment. SEPT preserves this semantic similarity, keeping strong alignment between each class and its neighbors.
    }
    \label{fig:motivation}
\end{figure}

Prompt tuning has emerged as an effective technique for adapting pretrained vision–language models (VLMs) such as CLIP~\citep{clip} to downstream tasks. 
Rather than relying on handcrafted templates, it learns continuous prompt embeddings optimized for the task. 
CoOp~\citep{coop} first showed that optimizing prompts on a small set of base classes can significantly improve performance. 
However, conventional prompt tuning often overfits to base (seen) classes, resulting in poor generalization to new (unseen) ones—a phenomenon known as the Base-New Tradeoff (BNT). 
BNT has become a central challenge in prompt tuning, motivating a wide range of solutions, including architectural enhancements~\citep{cocoop,dept,maple,clipadapter} and regularization strategies~\citep{kgcoop,prograd}. 
These efforts underscore the importance of learning prompts that generalize well across categories and domains.

The success of prompt tuning in VLMs has recently motivated its extension to audio–language models (ALMs)~\citep{clap22,clap23,laionclap,pengi}.
Audio-free prompt tuning methods~\citep{audiofree,clep_dg} have been proposed for sound event classification and emotion recognition. These approaches leverage task-specific textual descriptions of audio instead of raw audio inputs, aiming to improve generalization to unseen classes.
Meanwhile, PALM~\citep{palm} introduced a few-shot prompt tuning method for ALMs, showing strong performance across a range of audio classification tasks, including instrument, acoustic scene, and vocal sound classification.
Despite these advances, important generalization challenges—such as base-to-new generalization and cross-dataset transferability—remain insufficiently addressed in the context of prompt tuning for ALMs.

The generalization ability of VLMs and ALMs comes from a semantically well-structured text embedding space~\citep{frome2013devise, jia2021scaling, clip}.
When text embeddings of semantically related classes, such as words with similar meanings or usage contexts, are organized as coherent clusters, the model is better equipped to generalize to unseen classes.
We conduct a preliminary study to examine how prompt tuning in ALMs affects the structure of the text embedding space.
As shown in Fig.~\ref{fig:motivation}, hand-crafted zero-shot prompts (``This is a sound of \{class\}'') naturally preserve semantic structure: classes (y-axis) maintain strong similarity with their semantically related neighbors (x-axis), (e.g., synonyms or paraphrases), as indicated by the high similarity scores for both base and new classes.
In contrast, CoOp~\citep{coop} tends to disrupt this desirable structure when applied to ALMs.
Due to the overfitting of learned prompt embeddings only with seen class names, the similarity between a class and its semantic neighbor weakens, indicating that the semantic relationship is not adequately preserved. 
We attribute the severity of this vulnerability in ALMs to the inherent semantic sparsity of audio benchmarks~\citep{urbansound, vocalsound, crema-d}, which are constrained to only tens of classes.
In such a sparse semantic space, the learned prompts lack sufficient support to maintain geometric cohesion, degenerating into isolated prototypes rather than robust semantic anchors~\citep{harun2024variables, wu2018improving}.

To address the limitations of prompt overfitting and poor semantic alignment, we propose \textbf{S}emantically \textbf{E}xpanded \textbf{P}rompt \textbf{T}uning (\textbf{SEPT})—a novel and plug-and-play prompt tuning framework that explicitly regularizes the structure of the text embedding space by leveraging semantic neighbors.
Our approach utilizes the large language models to enrich the semantic coverage of each class.
For every seen class, we generate a diverse set of semantically related neighbor classes, and these neighbors are incorporated into the prompt tuning process to encourage each class and its semantic variants to form a coherent cluster in the shared text embedding space.
We explicitly enforce a robust geometric structure within this enriched space by introducing a semantic expansion loss that encourages the embedding of each class to be close to its own semantic neighbors (i.e., positive neighbors), while maintaining sufficient distance from the neighbors of all other classes (i.e., negative neighbors).
This pull–push mechanism promotes intra-class compactness and inter-class separability in the embedding space, thereby encouraging a semantically coherent structure that is critical for generalization to unseen classes.
In addition, to avoid over-compression of positives and excessive repulsion of negatives, we propose margin constraints for both intra-class and inter-class relationships.

Our proposed SEPT is model-agnostic and plug-and-play, and can be seamlessly integrated into a wide range of existing prompt tuning approaches.
To demonstrate its versatility, we apply our method to several representative baselines.
To the best of our knowledge, we are the first to establish a comprehensive generalization evaluation setup for ALMs and to rigorously evaluate prompt generalization across diverse datasets.
Through comprehensive experiments on multiple audio classification benchmarks, we demonstrate that our method consistently improves the performance of all these baselines, both base-to-new generalization and cross-dataset transferability.

\section{Related Work}
\label{sec:related_work}

\paragraph{Prompt Tuning in Vision-Language Models.}
Prompt tuning has emerged as a promising technique for vision-language models (VLMs), with the introduction of CLIP~\citep{clip}, which aligns vision and language modalities via contrastive learning.
To adapt CLIP to downstream tasks with minimal resources, CoOp~\citep{coop} proposed optimizing a set of learnable context tokens instead of relying on handcrafted textual prompts while keeping encoders frozen.
However, the learned context tokens tend to overfit the classes seen during training, which weakens the general knowledge of CLIP and leads to a significant performance drop on unseen data, which is also known as Base-New Tradeoff (BNT).

To alleviate this issue, CoCoOp~\citep{cocoop} introduced an input-conditional context token by leveraging an additional neural network called Meta-Net. 
Extending prompt tuning beyond the language branch, MaPLe~\citep{maple} jointly learns prompts for both vision and language encoders, with coupled prompt representations that explicitly model cross-modal interactions and improve generalization.
KgCoOp~\citep{kgcoop} addresses the problem 
by adding a regularization term that minimizes the Euclidean distance between learned prompt embeddings and handcrafted, knowledge-rich general prompt embeddings.
Similarly, LASP~\citep{lasp} also mitigates base-class overfitting by text-to-text objective that encourages soft prompts to remain close to handcrafted textual prompts.
DePT~\citep{dept} analyzes the BNT as a channel bias problem, proposing a decoupling strategy that isolates base-specific knowledge to preserve task-shared knowledge, significantly enhancing generalization to unseen tasks.
More recently, DPC \citep{dpc} further alleviates BNT by using two collaborating prompts, and TAC \citep{tac} improves generalization by injecting task-aware tokens derived from class clustering.
Collectively, these works demonstrate the efficacy of prompt tuning for few-shot generalization in VLMs.

\begin{figure*}[ht!]
    \centering
    \includegraphics[width=0.9\textwidth]{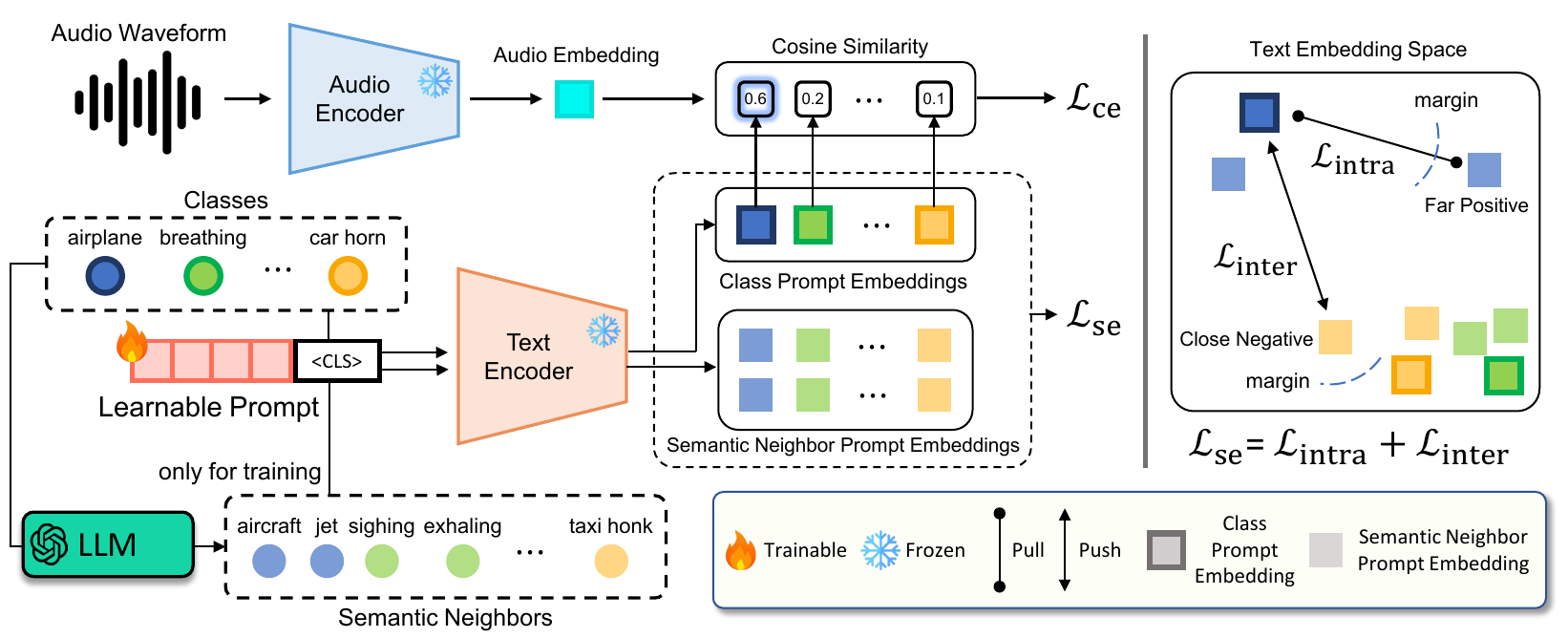}
    \caption{Framework of proposed Semantically Expanded Prompt Tuning (SEPT), applied to CoOp \cite{coop}. 
    During training, we use LLM to generate Semantic Neighbors-words that are semantically related to each class name.
    These neighbors are encoded into semantic neighbor embeddings using the same learnable prompt and frozen text encoder.
    To preserve the semantic structure of the original text embedding space, we introduce a semantic expansion loss $\mathcal{L}_{\text{se}}$.
    Specifically, as illustrated in the figure, $\mathcal{L}_{\text{intra}}$ pulls class embeddings closer to far semantic neighbors, and $\mathcal{L}_{\text{inter}}$ pushes away from overly close unrelated ones.
    Note that our method can be integrated into other prompt tuning methods and does not affect inference efficiency.
    }
    \label{fig:method}
\end{figure*}

\paragraph{Prompt Tuning in Audio-Language Models.}
The success of prompt tuning for VLMs has recently inspired its application in audio-language models (ALMs).
Test-time domain adaptation for ALMs~\citep{test_time_adaptation} adapts learnable prompts by learning domain representations directly from test audio, requiring neither extra annotations nor access to training data.
PT-Text~\citep{audiofree} and CLEP-DG~\citep{clep_dg} propose audio-free prompt tuning strategies, where learnable prompts are trained using task-specific audio descriptions instead of raw audio, thereby improving generalization beyond seen classes.
PALM~\citep{palm} explores few-shot prompt tuning by adapting learnable prompts in feature space, enhancing performance under limited supervision.
However, the generalization problem, which is a critical research focus of prompt tuning for VLMs, remains underexplored in the context of ALMs across diverse audio classification datasets.

\label{sec:method}
\section{Method}
We present Semantically Expanded Prompt Tuning (SEPT), a novel and plug-and-play framework for prompt tuning that enhances the structure of the text embedding space by incorporating semantic neighbors.
Our method explicitly regularizes the structure of the text embedding space to form semantically meaningful clusters, as shown in Fig.~\ref{fig:method}.

\subsection{Preliminaries} 
\label{preliminaries}

Audio-Language Models (ALMs)~\citep{clap22,clap23,laionclap,pengi} align audio and text modalities via contrastive learning on large-scale paired datasets. By optimizing contrastive loss, ALMs bring matching pairs closer in a shared embedding space while separating unrelated ones, effectively bridging the modality gap.

Zero-shot audio classification uses the pre-trained ALM without any additional fine-tuning on the target task. Let $\mathbf{x}$ be an audio embedding obtained from the audio encoder, and $\{\mathbf{w}_i\}_{i=1}^K$ be text embeddings generated from class-descriptive prompts (e.g., ``This is a sound of \{class\}'') by the text encoder, using the class $\{c_i\}_{i=1}^{K}$ as class name. The prediction probability is defined as:
\begin{equation}
\label{eq:pred_zs_audio}
p(y | \mathbf{x}) = \frac{\exp (\operatorname{sim} (\mathbf{x}, \mathbf{w}_y) / \tau)}{\sum_{i=1}^K \exp (\operatorname{sim} (\mathbf{x}, \mathbf{w}_i) / \tau)},
\end{equation}
where $\operatorname{sim}(\cdot, \cdot)$ is the cosine similarity and $\tau$ is a temperature parameter.

Prompt-based tuning replaces fixed hand-crafted templates with learnable prompts—continuous context vectors that are optimized alongside the downstream task objective. Specifically, for each class name $c_{i}$ with token embedding function $\mathcal{E}$, $M$ learnable context vectors
$[\mathbf{v}_1, \mathbf{v}_2, \dots, \mathbf{v}_M]$
are concatenated with token embedding $\mathcal{E}(c_i)$ to form the prompt:
{$\mathbf{t}_i = [\mathbf{v}_1, \dots, \mathbf{v}_M, \mathcal{E}({c}_i)]$}. These prompts are then passed to the frozen text encoder $g(\cdot)$ to produce the class prompt embedding. 
Given an audio input $\textbf{x}$ and class prompt embeddings $g(\textbf{t}_i)$, the prediction is made using:
\begin{equation} \label{eq:pred_coop_audio}
p(y | \mathbf{x}) = \frac{\exp (\operatorname{sim} (\mathbf{x}, g(\mathbf{t}_y)) / \tau)}{\sum_{i=1}^K \exp (\operatorname{sim} (\mathbf{x}, g(\mathbf{t}_i)) / \tau)},
\end{equation}
where $y$ is the corresponding label.
The prompt vectors are updated using a cross-entropy loss, while the audio and text encoders are kept frozen, thus maintaining the generalization capacity of the pre-trained ALM. The cross-entropy loss is given by:
\begin{equation}
\label{eq:loss_ce}
\mathcal{L}_{\mathrm{ce}} = - \sum_{\mathbf{x} \in \mathbf{X}} \log p(y|\mathbf{x}).
\end{equation}

\subsection{Semantic Neighbors}
Unlike image classification benchmarks with thousands of categories~\citep{imagenet}, most audio datasets contain only tens of classes~\citep{ns-instruments}, providing far narrower semantic coverage. 
This scarcity amplifies the overfitting issue observed in CoOp, causing class embeddings to become overly isolated and weakening their similarity to semantic neighbors, as illustrated in Fig.~\ref{fig:motivation}.
Such degradation of semantic structure makes it substantially more difficult for the model to generalize to unseen classes.

To address this, we introduce a semantic expansion strategy that leverages semantic neighbors. Specifically, for each class name $c_i$, we generate a set of $N$ semantically related neighbors $\{p_i^1,\dots,p_i^N\}$ using a large language model (LLM).
These neighbors capture fine-grained acoustic variants and natural language expressions related to $c_i$, effectively forming a dense semantic cluster around each original class in the text embedding space.
we structure the semantic expansion loss such that the prompt embedding for each class is pulled closer to its own neighbors (i.e., positive neighbors), while being pushed away from the neighbors of all other classes (i.e., negative neighbors). This dual mechanism encourages both intra-class compactness and inter-class separability, thereby expanding the robust semantic structure of the embedding space.

\subsection{Margin Constraint}
While pulling positive neighbors close and pushing negative neighbors away in embedding space can clarify the semantic structure, applying this mechanism naively—forcing all positives to be maximally close and all negatives maximally far—can paradoxically degrade the embedding space. Over-compressing positives suppresses the natural diversity of language and sound, while excessive separation of negatives can distort legitimate semantic relationships across classes.

To mitigate this, we introduce distance margin constraints $m_{i,j,n}$ that reflect their original semantic distance.
To ensure robustness, we compute this margin by averaging statistics across a diverse pool of \(T\) prompts \(\{\pi_t\}_{t=1}^T\) automatically generated using LLM (e.g., "a sound of", "a recording of").
The margin constraint between the $c_i$ and \(n\)-th neighbor of $c_j$, $p_{j}^{n}$, is defined as the average L₂ distance of their embeddings across all \(T\) prompts:
\begin{equation}
\label{eq:margin}
m_{i,j,n}
=
\frac{1}{T}
\sum_{t=1}^{T}
\Bigl\lVert\,
g\bigl(\mathcal{E}(\pi_{t};c_{i})\bigr)
-
g\bigl(\mathcal{E}(\pi_{t};p_{j}^{n})\bigr)
\Bigr\rVert_{2}.
\end{equation}
Here, \(g (\mathcal{E}(\pi_t;c_i))\) denotes the prompt embedding of the \(i\)-th class name \(c_i\) under prompt \(\pi_t\), \(g(\mathcal{E}(\pi_t;p_{j}^{n}))\) denotes the embedding of the \(n\)-th neighbor of class \(j\), \(p_{j}^{n}\), under the same prompt, and \(\lVert \cdot \rVert_{2}\) represents the Euclidean (L₂) norm.

Building on this margin, we introduce a novel semantic expansion loss that pulls class prompts towards positive neighbors and pushes them away from negative neighbors with a margin constraint.

\subsection{Semantic Expansion Loss}
We define two losses to regularize the prompt embedding space via semantic expansion: (i) an intra-class alignment loss $\mathcal{L}_{\mathrm{intra}}$, which encourages each class embedding to be close to its positive neighbors, and (ii) an inter-class separation loss $\mathcal{L}_{\mathrm{inter}}$, which pushes it away from negative neighbors.

As defined in the preliminaries, each class prompt embedding for the class $c_i$ is given by
\begin{equation}
\mathbf z_i = g(t_i) = g\bigl([\mathbf v_1, \mathbf v_2, \dots, \mathbf v_M, \mathcal{E}(c_i)]\bigr),
\end{equation}
where \(\{\mathbf v_m\}_{m=1}^M\) are the learnable context vectors and \(\mathcal{E}( c_i)\) is the token embedding of the class $c_{i}$.
We also define $\{\mathbf{p}_i^n\}_{n=1}^N$ as prompt embeddings of semantic neighbors of class $c_{i}$, which can be expressed as 
\begin{equation}
\mathbf p_i^{n} = g\bigl([\mathbf v_1, \mathbf v_2, \dots, \mathbf v_M, \mathcal{E}(p_i^{n})]\bigr).
\end{equation}

To encourage tighter semantic cohesion within each class, we pull the class prompt embedding $\mathbf z_i$ closer to its own neighbors $\mathbf{p}_i^n$ in the embedding space, if the distance is larger than the margin. The intra-class alignment loss is computed as a margin-based hinge loss:
\begin{equation}
\label{eq:intra}
\mathcal{L}_\text{intra}^{(i)} =
\frac{1}{N}
\sum_{n=1}^{N}
\max \big(0,\,
\lVert \mathbf{z}_i - \mathbf{p}_i^n \rVert_2 - m_{i,i,n}
\big),
\end{equation}
where \( m_{i,i,n} \) is the precomputed intra-class margin between embeddings of the $c_i$ and its neighbor $p_i^n$ as defined in Eq.~\ref{eq:margin}. 

Furthermore, to preserve inter-class distinctiveness, we push the class embedding $\mathbf z_i$ away from the neighbors $\{\mathbf{p}_j^n\}_{n=1}^N$ of all other classes $j \ne i$, by enforcing a minimum semantic distance via:
\begin{equation}
\label{eq:inter}
\mathcal{L}_\text{inter}^{(i,j)} =
\frac{1}{N}
\sum_{n=1}^{N}
\max \big(0,\,
m_{i,j,n} - \lVert \mathbf{z}_i - \mathbf{p}_j^n \rVert_2
\big),
\end{equation}
where \( m_{i,j,n} \) reflects the average margin distance between embeddings of $c_i$ and the $n$-th neighbor of class $j$, $p_j^n$.

The semantic expansion loss across all classes is computed by averaging intra- and inter-class losses:
\begin{equation}
\mathcal{L}_{\text{se}} =
\frac{1}{K} \sum_{i=1}^K
\left( \mathcal{L}_\text{intra}^{(i)}
+
\frac{1}{K-1} \sum_{\substack{j=1, j \ne i}}^K \mathcal{L}_\text{inter}^{(i,j)} \right).
\end{equation}
This loss encourages a structured expansion of the text embedding space, where each class forms a compact cluster of semantically related variants while remaining discriminative from other classes (e.g., “bell” and “chime” remain relatively close, while “explosion” and “chirp” are farther apart).
%

The total training loss combines the cross-entropy loss $\mathcal{L}_{\text{ce}}$ with the semantic expansion loss $\mathcal{L}_{\text{se}}$, enabling the model to learn from both supervised signals and semantic structure regularization: 
\[
\mathcal{L}_{\text{total}} = \mathcal{L}_{\text{ce}} + \lambda \cdot \mathcal{L}_{\text{se}},
\]
where $\lambda$ is a balancing hyperparameter.
Note that our $\mathcal{L}_{\text{se}}$ can be incorporated in a compatible manner alongside other auxiliary losses designed to enhance prompt tuning.

\begin{table*}[ht!]
  \centering 
  \footnotesize
  \scalebox{1}{
    \begin{tabular}{lcccccccccccc}
      \toprule
     \multicolumn{1}{l}{\multirow{2}{*}{\makecell[c]{{Method}}}}  & \multicolumn{3}{c}{Avg. over 11 datasets}  & \multicolumn{3}{c}{\cellcolor{blue!0}{Beijing-Opera}} & \multicolumn{3}{c}{{NS-Instruments}}  & \multicolumn{3}{c}{{ESC50}}  \\
    \cmidrule(lr){2-4}\cmidrule(lr){5-7}\cmidrule(lr){8-10}\cmidrule(lr){11-13}
    &  \multicolumn{1}{c}{Base} & \multicolumn{1}{c}{New} & \multicolumn{1}{c}{H} &  \multicolumn{1}{c}{Base} & \multicolumn{1}{c}{New} & \multicolumn{1}{c}{H}  & \multicolumn{1}{c}{Base} & \multicolumn{1}{c}{New} & \multicolumn{1}{c}{H}  & \multicolumn{1}{c}{Base} & \multicolumn{1}{c}{New} & \multicolumn{1}{c}{H} \\
    \midrule
        {Pengi}      & 40.68 & 38.21 & 38.46 & 64.16 & 68.55 & 66.15 & 43.32 & 30.01 & 35.46 & 15.70 & 13.70 & 14.33 \\
    \midrule
{CoOp}       & \textbf{65.00} & 34.09 & 42.83 & 97.27 & 61.38 & 74.87 & 41.11 & 23.88 & 30.08 & \textbf{61.97} & 11.83 & 19.46 \\
\rowcolor{gray!20}
\textbf{+ SEPT} & 64.36 & \textbf{42.98} & \textbf{49.70} & \textbf{97.88} & \textbf{71.87} & \textbf{82.45} & \textbf{43.33} & \textbf{37.78} & \textbf{40.29} & 59.00 & \textbf{15.80} & \textbf{24.70} \\
\midrule
{CoCoOp}     & \textbf{69.13} & 36.83 & 46.26 & \textbf{97.86} & 70.80 & \textbf{81.84} & 47.42 & \textbf{37.66} & \textbf{41.84} & \textbf{70.13} & 13.63 & 22.78 \\
\rowcolor{gray!20}
\textbf{+ SEPT} & 68.63 & \textbf{42.59} & \textbf{50.65} & 97.85 & \textbf{71.06} & 81.60 & \textbf{52.15} & 29.96 & 37.95 & 69.10 & \textbf{17.33} & \textbf{27.62} \\
\midrule
{KgCoOp}     & 37.99 & 37.42 & 36.39 & 67.26 & 61.40 & 62.96 & 39.80 & 39.27 & 39.11 & 14.10 & 10.33 & 11.76 \\
\rowcolor{gray!20}
\textbf{+ SEPT} & \textbf{58.92} & \textbf{45.28} & \textbf{49.79} & \textbf{94.44} & \textbf{67.99} & \textbf{78.28} & \textbf{51.24} & \textbf{40.86} & \textbf{45.28} & \textbf{47.60} & \textbf{18.33} & \textbf{26.33} \\
\midrule
{DePT}       & 63.86 & 39.91 & 46.79 & \textbf{97.26} & 62.25 & \textbf{75.32} & 43.75 & 28.21 & 34.01 & \textbf{60.80} & 14.00 & 22.53 \\
\rowcolor{gray!20}
\textbf{+ SEPT} & \textbf{64.57} & \textbf{41.63} & \textbf{49.06} & 96.32 & \textbf{62.74} & 75.00 & \textbf{47.66} & \textbf{37.90} & \textbf{41.60} & 56.80 & \textbf{16.03} & \textbf{24.91} \\

    \bottomrule
    \toprule
     \multicolumn{1}{l}{\multirow{2}{*}{\makecell[c]{{Method}}}}  & \multicolumn{3}{c}{{ESC50-Actions}}  & \multicolumn{3}{c}{{UrbanSound8k}} & \multicolumn{3}{c}{{CREMA-D}} & \multicolumn{3}{c}{{RAVDESS}}    \\
    \cmidrule(lr){2-4}\cmidrule(lr){5-7}\cmidrule(lr){8-10}\cmidrule(lr){11-13}
    &  \multicolumn{1}{c}{Base} & \multicolumn{1}{c}{New} & \multicolumn{1}{c}{H} &  \multicolumn{1}{c}{Base} & \multicolumn{1}{c}{New} & \multicolumn{1}{c}{H}  &  \multicolumn{1}{c}{Base} & \multicolumn{1}{c}{New} & \multicolumn{1}{c}{H}  &  \multicolumn{1}{c}{Base} & \multicolumn{1}{c}{New} & \multicolumn{1}{c}{H}  \\
     \midrule
     {Pengi}      & 25.50 & 28.00 & 26.44 & 30.68 & 39.51 & 34.14 & 52.25 & 32.36 & 39.96 & 25.38 & 32.16 & 28.37 \\
\midrule
{CoOp}       & \textbf{85.33} & 49.33 & 61.97 & \textbf{61.86} & 31.22 & 40.94 & 52.50 & 9.01 & 15.29 & 51.01 & 27.75 & 35.88 \\
\rowcolor{gray!20}
\textbf{+ SEPT} & 82.50 & \textbf{52.67} & \textbf{63.90} & 61.22 & \textbf{36.34} & \textbf{44.95} & \textbf{52.75} & \textbf{42.06} & \textbf{44.17} & \textbf{54.55} & \textbf{33.92} & \textbf{41.39} \\
\midrule
{CoCoOp}     & \textbf{89.17} & 41.50 & 56.15 & \textbf{64.25} & \textbf{30.77} & \textbf{41.05} & \textbf{56.80} & \textbf{26.07} & \textbf{35.62} & 56.82 & 26.87 & 36.39 \\
\rowcolor{gray!20}
\textbf{+ SEPT} & 87.50 & \textbf{51.50} & \textbf{64.58} & 61.06 & 25.93 & 36.26 & 54.06 & 19.72 & 28.21 & \textbf{63.55} & \textbf{37.57} & \textbf{46.77} \\
\midrule
{KgCoOp}     & 36.50 & 30.67 & 32.88 & 26.14 & 26.15 & 25.40 & 42.70 & 55.43 & 42.48 & 27.53 & 27.46 & 27.42 \\
\rowcolor{gray!20}
\textbf{+ SEPT} & \textbf{67.83} & \textbf{51.50} & \textbf{57.97} & \textbf{56.96} & \textbf{38.77} & \textbf{45.32} & \textbf{53.06} & \textbf{69.53} & \textbf{59.92} & \textbf{45.58} & \textbf{32.31} & \textbf{37.79} \\
\midrule
{DePT}       & \textbf{81.33} & 44.67 & 57.12 & \textbf{63.21} & \textbf{37.28} & \textbf{46.05} & \textbf{51.56} & 33.19 & 33.54 & 52.65 & \textbf{32.01} & \textbf{39.28} \\
\rowcolor{gray!20}
\textbf{+ SEPT} & 79.67 & \textbf{52.67} & \textbf{63.02} & 61.11 & 36.22 & 44.79 & 51.44 & \textbf{35.60} & \textbf{38.93} & \textbf{52.90} & 29.07 & 37.49 \\

    \bottomrule
    \toprule
     \multicolumn{1}{l}{\multirow{2}{*}{\makecell[c]{{Method}}}}  & \multicolumn{3}{c}{{SESA}}  & \multicolumn{3}{c}{{GT-Music-Genre}} & \multicolumn{3}{c}{{VocalSound}}  & \multicolumn{3}{c}{{TUT2017}}  \\
    \cmidrule(lr){2-4}\cmidrule(lr){5-7}\cmidrule(lr){8-10}\cmidrule(lr){11-13}
    &  \multicolumn{1}{c}{Base} & \multicolumn{1}{c}{New} & \multicolumn{1}{c}{H} &  \multicolumn{1}{c}{Base} & \multicolumn{1}{c}{New} & \multicolumn{1}{c}{H}  &  \multicolumn{1}{c}{Base} & \multicolumn{1}{c}{New} & \multicolumn{1}{c}{H}  &  \multicolumn{1}{c}{Base} & \multicolumn{1}{c}{New} & \multicolumn{1}{c}{H}   \\
    \midrule
    {Pengi}      & 75.00 & 89.19 & 81.48 & 35.64 & 15.15 & 21.26 & 59.28 & 53.01 & 55.97 & 20.52 & 18.65 & 19.53 \\
    \midrule
{CoOp}       & \textbf{85.78} & 74.77           & 79.33           & \textbf{57.76} & 20.54           & 29.76           & \textbf{75.26} & 45.01           & 56.00           & 45.11           & 20.25           & 27.53           \\
\rowcolor{gray!20}
\textbf{+ SEPT} & 78.92           & \textbf{83.78}  & \textbf{81.22}  & 56.11          & \textbf{23.23}  & \textbf{32.15}  & 74.11          & \textbf{54.14}  & \textbf{62.38}  & \textbf{47.55}  & \textbf{21.14}  & \textbf{29.05}  \\
\midrule
{CoCoOp}     & 85.78           & 72.97           & 78.68           & \textbf{60.73} & 24.24           & 32.68           & \textbf{82.47}          & 40.03           & 53.19           & 49.01           & 20.55           & 28.61           \\
\rowcolor{gray!20}
\textbf{+ SEPT} & \textbf{88.73}  & \textbf{86.49}  & \textbf{87.42}  & 49.80          & \textbf{46.32}  & \textbf{47.99}  & 79.57          & \textbf{61.06}  & \textbf{68.84}  & \textbf{51.60}  & \textbf{21.54}  & \textbf{29.94}  \\
\midrule
{KgCoOp}     & 68.14           & 79.28           & 72.37           & 27.39          & 15.82           & 19.77           & 52.50          & 49.91           & 51.10           & 15.83           & 15.90           & 15.10           \\
\rowcolor{gray!20}
\textbf{+ SEPT} & \textbf{79.90}  & \textbf{82.88}  & \textbf{81.23}  & \textbf{42.90} & \textbf{20.88}  & \textbf{28.08}  & \textbf{70.12} & \textbf{56.05}  & \textbf{62.23}  & \textbf{38.48}  & \textbf{18.94}  & \textbf{25.30}  \\
\midrule
{DePT}       & 80.39           & \textbf{91.89}  & \textbf{85.72}  & 50.17          & 22.90           & 31.08           & 72.24          & 49.00           & 58.28           & \textbf{49.12}  & \textbf{23.67}  & \textbf{31.75}  \\
\rowcolor{gray!20}
\textbf{+ SEPT} & \textbf{87.25}  & 82.88           & 84.90           & \textbf{55.78} & \textbf{23.91}  & \textbf{33.13}  & \textbf{75.34} & \textbf{57.70}  & \textbf{65.24}  & 46.05           & 23.25           & 30.70           \\

    \bottomrule
    \end{tabular}
}
\caption{Comparison on the base-to-new
generalization. Prompts are learned using 16-shot samples per class. 'Base' reports accuracy on seen classes, 'New' on unseen classes, and 'H' is the harmonic mean.
}
\label{tab:b2n_main}
\end{table*}

\label{sec:experiments}
\section{Experiments}

\subsection{Experimental Setup}

\begin{table}[ht!]
  \centering
  
  \label{tab:transfer}
  \scalebox{0.62}{
  \begin{tabular}{lcccccc}
    \toprule
\multirow{3}{*}{Method} 
  & \multicolumn{2}{c}{Sound Event Classif.} 
  & \multicolumn{2}{c}{Emotion Recog.} 
  & \multicolumn{2}{c}{Instrument Classif.} \\
\cmidrule(lr){2-3} \cmidrule(lr){4-5} \cmidrule(lr){6-7}
  & \makecell{ESC50-A.\\(Source)} & \makecell{UrbanS.\\(Target)} 
  & \makecell{RAV.\\(Source)}      & \makecell{CREM.\\(Target)} 
  & \makecell{NS-Inst.\\(Source)} & \makecell{Beijing.\\(Target)} \\

    \midrule
CoOp         & \textbf{70.08} & 19.02           & 30.68           & 15.74           & 35.30           & 31.92           \\
\rowcolor{gray!20}
\textbf{+ SEPT}      & 69.42           & \textbf{24.21}  & \textbf{31.70}  & \textbf{23.89}  & \textbf{37.21}  & \textbf{41.66}  \\
\midrule
CoCoOp       & 77.83           & 16.43           & \textbf{33.81}  & 8.98            & 37.45           & 36.44           \\
\rowcolor{gray!20}
\textbf{+ SEPT}      & \textbf{78.17}  & \textbf{23.18}  & 31.50           & \textbf{19.14}  & \textbf{39.63}  & \textbf{37.99}  \\
\midrule
KgCoOp       & 15.92           & 13.34           & 14.19           & 14.08           & 16.55           & 29.96           \\
\rowcolor{gray!20}
\textbf{+ SEPT}      & \textbf{54.50}  & \textbf{21.84}  & \textbf{25.05}  & \textbf{20.60}  & \textbf{39.16}  & \textbf{50.01}  \\
\midrule
DePT         & \textbf{70.83}  & 21.86           & \textbf{30.96}  & 17.53           & \textbf{36.89}  & 37.72           \\
\rowcolor{gray!20}
\textbf{+ SEPT}      & 68.08           & \textbf{24.46}  & 27.83           & \textbf{27.07}  & 36.30           & \textbf{44.49}  \\

    \bottomrule
  \end{tabular}
  }
    \caption{Comparison on cross‐dataset evaluation. Prompts are trained only on the source dataset with 16-shot samples per class, and then directly applied to both the source and a different target dataset. Reported values are accuracy (\%).
    }

\label{tab:cross_dataset_main}
\end{table}

\paragraph{Datasets.}
Following PALM~\citep{palm}, we evaluate our method on a diverse collection of audio classification datasets, covering a wide range of speech and non-speech domains. 
We used Beijing-Opera~\citep{beijing-opera} and NS-Instruments~\citep{ns-instruments} for instrument classification, ESC50~\citep{esc50}, ESC50-Actions, and UrbanSound8K~\citep{urbansound} for sound event classification, CREMA-D~\citep{crema-d} and RAVDESS~\citep{ravdess} for emotion recognition, SESA~\citep{sesa} for surveillance sound classification, TUT2017~\citep{tut2017} for acoustic scene classification, GT-Music-Genre~\citep{gt-music-genre} for music analysis, and VocalSound~\citep{vocalsound} for vocal sound classification.

\paragraph{Model and Baselines.}
We utilize the audio and text encoders of the Pengi~\citep{pengi} for audio classification tasks following PALM. 
Although Pengi was originally designed for multimodal generation tasks, we adopt its audio and text encoders for classification, similar to prior work in prompt tuning. 
To demonstrate the compatibility of SEPT with existing prompt tuning methods, we apply it to several representative textual prompt tuning methods: CoOp~\citep{coop}, CoCoOp~\citep{cocoop}, KgCoOp~\citep{kgcoop}, and DePT~\citep{dept}, all of which are built upon CoOp.
Note that SEPT is designed as a model-agnostic and plug-and-play module and can be seamlessly integrated into other ALM or prompt tuning methods as well.

\paragraph{Implementation Details.}
For training, we randomly sample 16 instances per class from the training set to learn generalizable prompts, while keeping both the audio and text encoders frozen.
The number of semantic neighbors per class $N$ is set to 10, and the number of hand-crafted prompts used for margin computation $T$ is 100.
We employ ChatGPT-4o as the primary LLM in all experiments.
The balancing hyperparameter $\lambda$ is set to 3 by default.
All results are reported as the average over three random seeds.
Additional experimental details, including the prompts for generating semantic anchors and the hand-crafted prompt pool constructed by LLM, are provided in the Appendix~\ref{appndix:implementation_details}.

\subsection{Comparative Study}

\paragraph{Base-to-New Generalization.}
The base-to-new generalization setting evaluates a model's ability to transfer knowledge from base classes to previously unseen classes, reflecting a category shift between training and evaluation.
Following the previous works on visual-language prompt tuning~\citep{cocoop, kgcoop}, we divide the classes of each dataset equally into two splits, designated as base and new classes.
To evaluate the generalization ability of the prompt tuning methods, all the compared methods are trained using only the base classes.
During inference, we measure the accuracy separately for the base and new class splits, and report the harmonic mean (H) of these two accuracies as the primary evaluation metric.
 
Table~\ref{tab:b2n_main} presents a detailed comparison across 11 diverse audio classification benchmarks. We evaluate four representative prompt tuning—CoOp, CoCoOp, KgCoOp, and DePT—both with and without our SEPT.
Across all datasets and methods, incorporating our semantic expansion strategy markedly improves the harmonic mean and new class accuracy, while maintaining competitive performance on base classes.
In particular, CoOp and CoCoOp without our approach achieve high accuracy on base classes, but suffer from significant drops on new classes, reflecting the well-known base-to-new trade-off. By integrating SEPT, new class performance increases substantially (e.g., CoOp: 34.09\% $\to$ 42.98\%, CoCoOp: 36.83\% $\to$ 42.59\% on average), resulting in clear gains in the harmonic mean. 
Collectively, results demonstrate that our framework can serve as a compatible module for previous prompt tuning approaches, consistently improving base-to-new generalization and mitigating the overfitting in base classes commonly observed in conventional prompt optimization for audio classification tasks.
Note that the comparatively lower performance of KgCoOp in ALM, in contrast to VLM, is analyzed in detail in Appendix~\ref{appendix:kgcoop_analysis}.

\paragraph{Cross-Dataset Evaluation.}
To verify transferability beyond category shift and address dataset shift, we conducted cross-dataset evaluations. In this setup, prompts are trained on a source dataset and deployed for inference on a different target dataset without additional fine-tuning.
 
As summarized in Table~\ref{tab:cross_dataset_main}, SEPT substantially improves cross-dataset generalization across diverse audio classification tasks, including sound event classification, emotion recognition, and instrument recognition.
Baseline methods such as CoOp, CoCoOp, and DePT achieve decent performance on their source datasets but suffer notable accuracy drops when transferred to unseen target datasets.
In contrast, incorporating our semantic expansion strategy yields significant gains in transfer accuracy while maintaining competitive source-domain performance. These results demonstrate that our semantic expansion loss serves as a versatile, plug-and-play regularizer that enhances the robustness of the prompt space, which is crucial for successful cross-dataset transfer in ALMs.

\begin{table}[t!]
\centering
\scalebox{0.7}{%
\begin{tabular}{cccccccc}
\toprule
  & $\mathcal{L}_\text{intra}$ & $m^\text{intra}$ & $\mathcal{L}_\text{inter}$ & $m^\text{inter}$ & Base & New & H \\
\midrule
1 &             &             &            &            & 65.00 & 34.09 & 42.83 \\
2 & \checkmark  &             &            &            & 58.31 & 33.58 & 41.17 \\
3 & \checkmark  & \checkmark  &            &            & 64.99 & 36.90 & 44.52 \\
4 &             &             & \checkmark &            & 62.01 & 36.68 & 43.48 \\
5 &             &             & \checkmark & \checkmark & 62.84 & 38.79 & 46.04 \\
6 & \checkmark  &             & \checkmark &            & 61.82 & 32.89 & 41.47 \\
\rowcolor{gray!20}
7 & \checkmark  & \checkmark  & \checkmark & \checkmark & 64.37 & 42.93 & \textbf{49.70} \\
\bottomrule
\end{tabular}%
}
\caption{Ablation of semantic expansion and margins. $m^\text{intra}$ and $m^\text{inter}$ indicate whether a margin constraint is applied to $\mathcal{L}_\text{intra}$ and $\mathcal{L}_\text{inter}$, respectively.}
\label{tab:ablation_flags}
\end{table}

\subsection{Analysis}
\label{sec:analysis}

\paragraph{Semantic Expansion with Margin Constraints.}
\textit{}Table~\ref{tab:ablation_flags} presents an ablation study on the semantic expansion components of SEPT built on CoOp, specifically the intra- and inter-class losses ($\mathcal{L}_\text{intra}$, $\mathcal{L}_\text{inter}$) and their margin constraints ($m^\text{intra}$, $m^\text{inter}$). $m^\text{intra}$ and $m^\text{inter}$ indicate whether margin constraints are applied to $\mathcal{L}_\text{intra}$ in Eq.~\ref{eq:intra} and $\mathcal{L}_\text{inter}$ in Eq.~\ref{eq:inter}, respectively. The first row reports baseline CoOp performance without semantic expansion.
When applying naive expansion losses without margin constraints (rows 2, 4, 6), performance does not improve over the baseline and often deteriorates on both base and new classes.  This shows that indiscriminately pulling prompts towards positives or pushing them away from negatives—without considering their semantic relationships—disrupts the embedding space, leading to over-compression or fragmentation of class clusters.
In contrast, introducing margin constraints (rows 3, 5, 7) substantially mitigates these issues. 
In particular, the full model with both intra- and inter-class margins (row 7) improves the harmonic mean with a large margin, from 42.83 to 49.70, demonstrating a clear gain in generalization.
These results show that margin constraints play a critical role in structuring the embedding space by controlling intra-class compactness and inter-class distinctiveness.

\begin{table}[t!]
\centering
\scalebox{0.65}{
\begin{tabular}{lcccc}
\toprule
            & \# Params & Training Time (s) & Inference Time (ms) & H \\
\midrule
CoCoOp      & 107,072   & 681.5     & 40.9     & 46.26 \\
KgCoOp      & 8,192     & 96.0     & 2.8      & 36.39 \\
DePT        & 114,738   & 108.1     & 2.8      & 46.79 \\
\cmidrule(lr){1-5}
CoOp        & 8,192     & 94.0     & 2.8      & 42.83 \\
\rowcolor{gray!20}
\textbf{+ SEPT}     & 8,192     & 170.7     & 2.8      & \textbf{49.70} \\
\bottomrule
\end{tabular}
}
\caption{Comparison of computational costs. \# Params means the number of trainable parameters. The training time is calculated for 50 epochs, while the inference time is measured as the time taken per sample.
}
\label{tab:complexity}
\end{table}

\paragraph{Complexity.}

We evaluate the computational efficiency of SEPT by comparing the number of trainable parameters, training$/$inference time against baseline methods on the ESC50 dataset (Table~\ref{tab:complexity}).
Crucially, SEPT operates as a plug-and-play module that introduces zero additional parameters and no inference latency overhead, maintaining the exact model size and high inference speed of the CoOp.
While the incorporation of semantic expansion incurs an increase in training time compared to CoOp, it remains significantly more efficient than architecture-heavy methods such as CoCoOp.
Ultimately, SEPT achieves a superior trade-off, delivering the highest generalization performance without compromising inference efficiency.

\begin{figure}[t!]
  \centering
  \begin{minipage}[b]{0.49\linewidth}
    \centering
    \includegraphics[width=\linewidth]{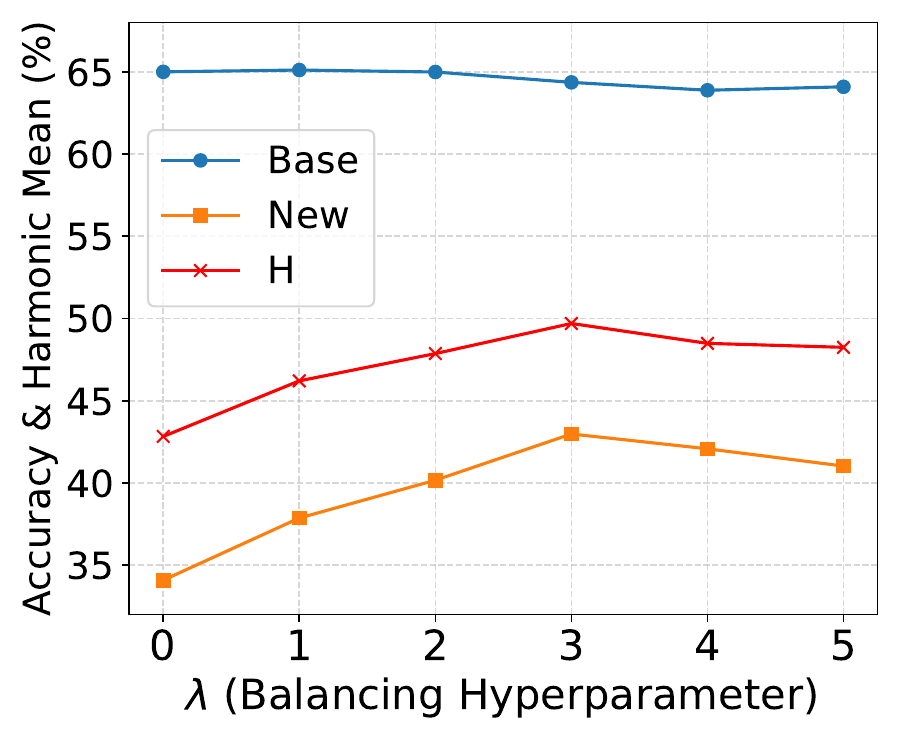}
  \end{minipage}
  \hfill
  \begin{minipage}[b]{0.49\linewidth}
    \centering
        \includegraphics[width=\linewidth]{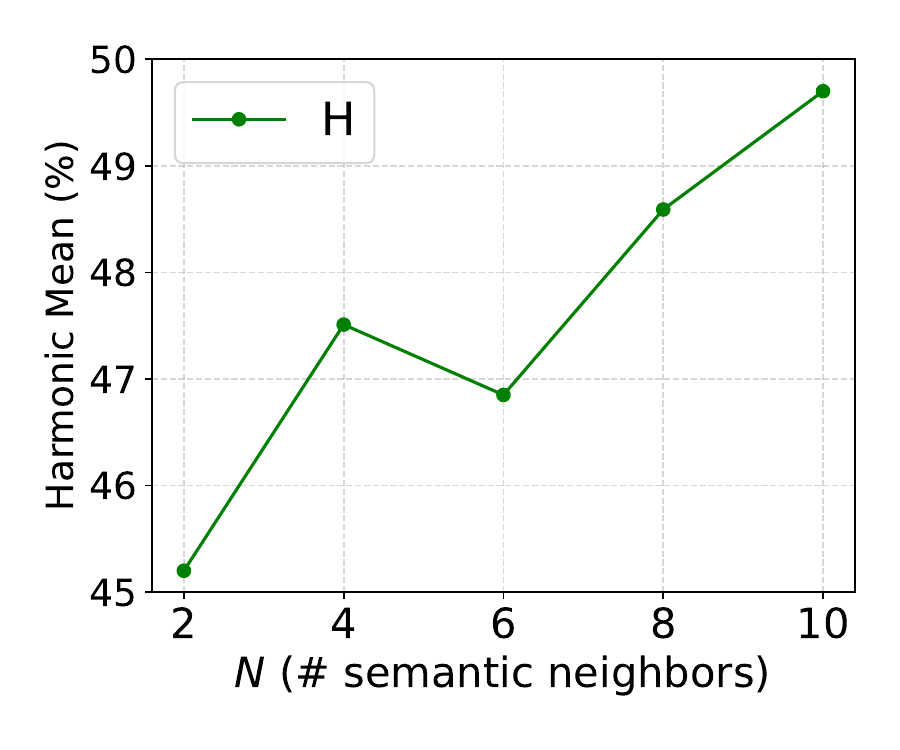}
  \end{minipage}

  \caption{Ablation studies of the hyperparameter $\lambda$ (left) and the number of semantic neighbors $N$ (right).}
  \label{fig:ablation_lambda_n_pseudo}
\end{figure}

\paragraph{Hyperparameter $\lambda$ and $N$.}
Figure~\ref{fig:ablation_lambda_n_pseudo} presents the ablation studies of two hyperparameters: the balancing parameter $\lambda$ and the number of semantic neighbors $N$.
Optimal performance is achieved at $\lambda=3$, which effectively balances supervised classification and structural regularization, yielding the highest harmonic mean.
Moreover, SEPT consistently outperforms the baseline across a wide range of $\lambda$ values, demonstrating its robustness to hyperparameter variations.
Meanwhile, increasing the number of semantic neighbors $N$ tends to improve the harmonic mean.
This indicates that incorporating a more diverse set of semantic neighbors provides richer semantic information, allowing the prompt embedding space to be better structured, boosting generalization performance.

\begin{figure}[t!]
  \centering
    \includegraphics[width=0.95\linewidth]{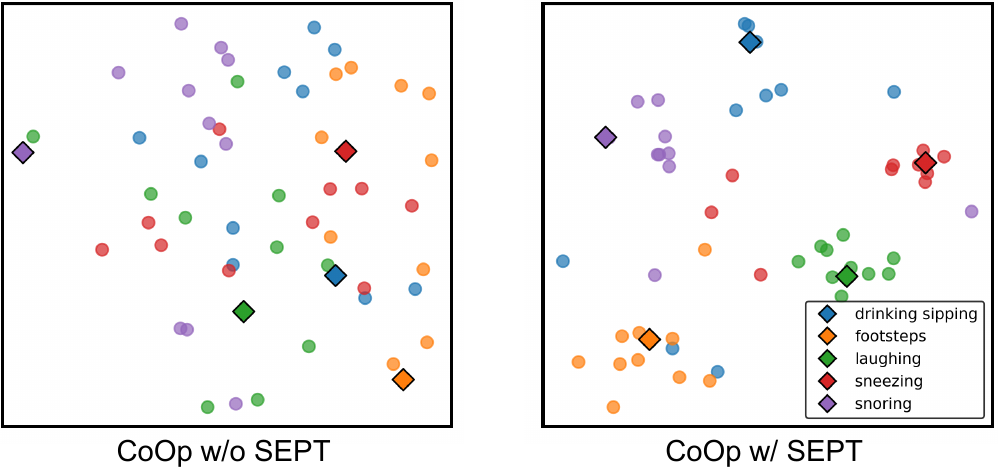}
  \caption{t-SNE visualization of prompt embeddings of new classes and their semantic neighbors from ESC50-Actions.
$\diamond$ denotes the prompt embedding of a new class, while $\circ$ represents its corresponding semantic neighbors.
  }
  
  \label{fig:tsne}
\end{figure}

\paragraph{Qualitative Analysis.}
To qualitatively assess whether semantic expansion with base class leads to better structured representations for new (unseen) classes, we visualize the learned text embeddings for new classes and their semantic neighbors using t-SNE, as shown in Fig.~\ref{fig:tsne}.
The scatterplots compare CoOp with and without our proposed SEPT framework on the ESC50-Actions.
Without SEPT (left), the text embedding space lacks structure—semantic neighbors are scattered and do not form coherent clusters. In contrast, with SEPT (right), each new class forms a compact cluster with its corresponding semantic neighbors, illustrating improved intra-class alignment and inter-class separation.
This result suggests that the semantic structure induced during training transfers effectively to unseen classes. Additional qualitative examples and neighbor lists are provided in Appendix~\ref{appndix:additional_qualitative}.

\subsection{Robustness Across Different LLMs.}
\label{sec:llm_dependency}
\begin{table}[t!]
\centering
\scalebox{0.8}{
\begin{tabular}{lccc}
\toprule
\multicolumn{1}{l}{\multirow{2}{*}{{Method}}} & \multicolumn{3}{c}{Avg over 11 datasets} \\
\cmidrule(l){2-4} 
 & {Base} & {New} & {H} \\ 
\midrule
CoOp                        & 65.00 & 34.09 & 42.83 \\
\midrule
+ \textbf{SEPT} (Qwen3 Max)          & 64.70 & 40.19 & 47.71 \\
+ \textbf{SEPT} (Gemini 2.5 Pro)     & 64.61 & 41.26 & 48.32 \\
+ \textbf{SEPT} (ChatGPT-4o)         & 64.36 & 42.98 & {49.70} \\
\bottomrule
\end{tabular}
}
\caption{Comparison of different LLMs for neighbor generation when applying SEPT to CoOp.}
\label{tab:llm_backbone_ablation}
\end{table}
We verify the robustness of our approach by employing various LLMs for neighbor generation. As shown in Table~\ref{tab:llm_backbone_ablation}, SEPT consistently achieves substantial gains over the CoOp baseline, regardless of the underlying LLM backbone. Notably, even with Qwen3 Max~\cite{qwen3} and Gemini 2.5 Pro~\cite{gemini}, our method yields impressive H-scores of 47.71 and 48.32, respectively. This demonstrates that the effectiveness of SEPT stems from the proposed semantic expansion strategy itself, rather than being dependent on a specific language model.

\section{Conclusion}
\label{sec:conclusion}
In this paper, we identify that prompt tuning in Audio-Language Models (ALMs) also suffers from the Base–New Tradeoff and disrupts the semantic structure of the prompt embedding space, leading to poor generalization to unseen classes.
To address this, we propose Semantically Expanded Prompt Tuning (SEPT), a simple yet effective framework designed to enhance the generalization capability of prompt tuning in ALMs.
SEPT leverages large language models to generate semantically related neighbors for each class, thereby enriching the prompt embedding space with meaningful semantic structure.
Concretely, we introduce a semantic expansion loss with margin constraints that jointly promote intra-class compactness and inter-class separability.
To rigorously evaluate SEPT, we introduce the first evaluation protocols for generalization in ALM prompt tuning.
Extensive experiments demonstrate that SEPT consistently improves existing methods in both base-to-new generalization and cross-dataset transferability.
Results highlight the versatility and effectiveness of SEPT as a plug-and-play module, paving the way for more robust and generalizable prompt tuning in ALMs.

\section{Limitations}

While the proposed SEPT framework achieves significant improvements by explicitly regularizing the textual prompt space, its underlying principle of semantic expansion holds potential for broader multimodal adaptation.
Currently, our implementation focuses on text-modality regularization to address the Base-New Trade-off. Given the recent emergence of learnable multimodal prompts as a powerful technique for downstream tasks, a promising avenue for future research is to extend SEPT to the audio prompt domain. Adapting our semantic expansion strategy to audio prompts could foster semantically enriched joint audio--text embeddings, potentially unlocking further generalization capabilities in fully multimodal ALM setups.

\section*{Acknowledgments}

This work was supported by the Institute of Information \& Communications Technology Planning \& Evaluation (IITP) grant funded by the Korea government (MSIT) (No. RS-2020-II200153, Penetration Security Testing of ML Model Vulnerabilities and Defense).
\bibliography{custom}

\newpage
\clearpage

\appendix
\label{sec:appendix}
\section*{\centering\LARGE Appendix}
\startcontents[appendixtoc]
\printcontents[appendixtoc]{l}{1}{\setcounter{tocdepth}{2}}
\addtocontents{toc}{\protect\setcounter{tocdepth}{2}}
\definecolor{linkcolor}{HTML}{000000}
\newpage

\definecolor{linkcolor}{HTML}{ED1C24}

\section{Detailed Experimental Settings}

\subsection{Dataset overview.}
Table~\ref{tab:base_new_classes} summarizes the information about the datasets used in this study. Following PALM~\citep{palm}, we evaluate prompt tuning methods on various downstream tasks, including instrument classification, sound event classification, emotion recognition, vocal sound classification, surveillance sound classification, acoustic scene classification, and music analysis.
For datasets with multiple folds, we conduct experiments on all folds, and the average accuracy across them is reported.
Each dataset is split into base and new classes with equal proportions, except for TUT2017~\citep{tut2017}, which consists of 15 classes divided into eight base classes and seven new classes.

\subsection{Implementation Details.}
\label{appndix:implementation_details}
All learnable prompts were trained for 50 epochs using a learning rate of 0.0125 and the SGD optimizer with a momentum of 0.9. In all experiments, the length of the learnable context vector $M$ was set to 16, with each vector having a dimensionality of 512.
Since no separate validation set was used, performance was evaluated using the model checkpoint with the lowest training loss.
All reported base accuracy, new accuracy, and harmonic mean (H) values are averaged over three random runs. For a fair comparison, DePT was implemented without test-time knowledge fusion.
Additionally, we excluded the batch normalization layers of the pretrained Pengi audio encoder during our experiments.
All experiments were conducted on a machine equipped with an Intel(R) Xeon(R) Silver 4210R CPU @ 2.40GHz and a single NVIDIA RTX 3090 GPU.
To compute a robust margin, we generate diverse prefixes using the prompts in~Fig. \ref{fig:appendix_prompt_prefix}.
The prompt for generating semantic neighbors for semantic expansion is in~Fig. \ref{fig:appendix_prompt_neighbors}.

\begin{table*}[t!]
\centering
\scalebox{0.6}{%
\begin{tabular}{llp{5.5cm}p{5.5cm}cc}
\toprule
\textbf{Task} & \textbf{Dataset} & \textbf{Base Classes} & \textbf{New Classes} & \textbf{\# Classes} & \textbf{Split} \\
\midrule
\multirow{2}{*}{Instrument Classification} 
& Beijing-Opera & bangu, daluo & naobo, xiaoluo & 4 & Five Fold \\
\cmidrule{2-6}
& NS-Instruments & bass, brass, flute, guitar, keyboard & mallet, organ, reed, string, vocal & 10 & Train–Test \\
\midrule
\multirow{13}{*}{Sound Event Classification} 
& ESC50-Actions & breathing, brushing teeth, clapping, coughing, crying baby & drinking sipping, footsteps, laughing, sneezing, snoring & 10 & Five Fold \\
\cmidrule{2-6}
& ESC50 & airplane, breathing, brushing teeth, can opening, car horn, cat, chainsaw, chirping birds, church bells, clapping, clock alarm, clock tick, coughing, cow, crackling fire, crickets, crow, crying baby, dog, door wood creaks, door wood knock, drinking sipping, engine, fireworks, footsteps & frog, glass breaking, hand saw, helicopter, hen, insects, keyboard typing, laughing, mouse click, pig, pouring water, rain, rooster, sea waves, sheep, siren, sneezing, snoring, thunderstorm, toilet flush, train, vacuum cleaner, washing machine, water drops, wind & 50 & Five Fold \\
\cmidrule{2-6}
& UrbanSound8K & air conditioner, car horn, children playing, dog bark, drilling & engine idling, gun shot, jackhammer, siren, street music & 10 & Ten Fold \\
\midrule
\multirow{2}{*}{Emotion Recognition} 
& CREMA-D & anger, disgust, fear & happy, neutral, sad & 6 & Train–Test \\
\cmidrule{2-6}
& RAVDESS & angry, calm, disgust, fearful & happy, neutral, sad, surprised & 8 & Train–Test \\
\midrule
Vocal Sound Classification 
& VocalSound & Cough, Laughter, Sigh & Sneeze, Sniff, Throat clearing & 6 & Train–Test \\
\midrule
Surveillance Sound Classification 
& SESA & casual, explosion & gunshot, siren & 4 & Train–Test \\
\midrule
Acoustic Scene Classification 
& TUT2017 & beach, bus, cafe/restaurant, car, city center, forest path, grocery store, home & library, metro station, office, park, residential area, train, tram & 15 & Four Fold \\
\midrule
Music Analysis 
& GT-Music-Genre & blues, classical, country, disco, hiphop & jazz, metal, pop, reggae, rock & 10 & Train–Test \\
\bottomrule
\end{tabular}%
}
\caption{Summary of datasets, including task, class split, and number of classes.}
\label{tab:base_new_classes}
\end{table*}

\begin{figure*}[ht!]
    \centering
    \begin{tcolorbox}[colback=gray!10, colframe=gray!50!black, boxrule=1pt, arc=2pt, width=1\textwidth, fontupper=\small]
    I am working on an audio classification task using CLAP (Contrastive Language-Audio Pretraining). In this setup, I embed both the audio signal and a text template like "a recording of" and then compute cosine similarity to classify the audio.
    
    However, I suspect that the default template "a recording of" may not be optimal for all audio types. For example, "a sound of", "the noise made by", or "an audio clip of" could potentially lead to better representations depending on the nature of the sound.
    
    I want to explore a diverse and creative list of candidate text templates that describe sounds, voices, audio events, or audio sources. Please generate at least 100 diverse, natural-sounding prompt templates without duplication that could plausibly describe a sound class, where <class> is the audio class label (e.g., dog barking, thunder, doorbell, car,  air conditioner, blues, etc.).
    
    Please vary:
    
    - the descriptive style (e.g., technical, casual, poetic)
    
    - the use of verbs and modifiers (e.g., “produced by,” “emitted from,” “generated by”)
    \end{tcolorbox}
    \caption{Prompt for diverse prefixes.}
    \label{fig:appendix_prompt_prefix}
\end{figure*}

\begin{figure*}[ht!]
    \centering
    \begin{tcolorbox}[colback=gray!10, colframe=gray!50!black, boxrule=1pt, arc=2pt, width=1\textwidth, fontupper=\small]
For each class name, I would like you to generate a list of 10 semantically similar terms that are specifically relevant in the context of audio, sound, or voice characteristics.

The output must adhere to the following criteria:

1. Each set of suggestions should be highly specific to the corresponding class name, reflecting similarity in terms of acoustic properties, usage contexts, or auditory perception.

2. The suggestions for each class must be semantically distinct from those of the other classes in the list, so that the resulting sets are clearly differentiated and non-overlapping.

3. Within each set, the suggestions should be diverse yet representative of the broader category or concept that unites them, ensuring a well-rounded and meaningful expansion.

Before generating the similar terms, please briefly analyze the given class list to identify any overarching themes or semantic groupings, and use this analysis to ensure the relevance and distinctiveness of the expanded sets.

The output should be returned in the following Python dictionary format:

{
'word1':[suggestion1, suggestion2,...],

'word2':[suggestion1, suggestion2,...],
}

Please ensure that the suggestions for each word reflect its unique role within the shared conceptual domain, and the suggested words for each provided word must not include any of the other words from the original list.
\end{tcolorbox}
    \caption{Prompts for semantic neighbors.}
    \label{fig:appendix_prompt_neighbors}
\end{figure*}

\begin{table*}[t]
\centering
\scalebox{0.9}{
\begin{tabular}{lcccccccccccc}
\toprule
\multirow{2}{*}{Method} & \multicolumn{3}{c}{\# shots = 2} & \multicolumn{3}{c}{\# shots = 4} & \multicolumn{3}{c}{\# shots = 8} & \multicolumn{3}{c}{\# shots = 16} \\
\cmidrule(lr){2-4}\cmidrule(lr){5-7}\cmidrule(lr){8-10}\cmidrule(lr){11-13}
 & Base & New & H & Base & New & H & Base & New & H & Base & New & H \\
\midrule
CoOp         & 47.95 & 33.29 & 37.69 & \textbf{53.98} & 33.95 & 39.34 & \textbf{60.62} & 36.28 & 43.70 & \textbf{65.00} & 34.09 & 42.83 \\
\rowcolor{gray!20}
\textbf{+ SEPT}      & \textbf{50.89} & \textbf{37.94} & \textbf{41.83} & 52.45 & \textbf{37.82} & \textbf{42.26} & 60.48 & \textbf{40.12} & \textbf{46.27} & 64.36 & \textbf{42.98} & \textbf{49.70} \\
\midrule
CoCoOp       & 50.91 & 38.10 & 42.06 & 54.73 & 36.55 & 41.96 & 62.30 & 37.94 & 45.02 & \textbf{69.13} & 36.83 & 46.26 \\
\rowcolor{gray!20}
\textbf{+ SEPT}      & \textbf{54.18} & \textbf{41.11} & \textbf{45.04} & \textbf{56.40} & \textbf{41.99} & \textbf{46.30} & \textbf{63.38} & \textbf{39.90} & \textbf{46.96} & 68.63 & \textbf{42.59} & \textbf{50.65} \\
\midrule
KgCoOp       & 35.24 & 37.69 & 35.22 & 35.31 & 38.76 & 35.75 & 36.17 & 38.33 & 36.08 & 37.99 & 37.42 & 36.39 \\
\rowcolor{gray!20}
\textbf{+ SEPT}      & \textbf{50.92} & \textbf{41.49} & \textbf{44.65} & \textbf{54.19} & \textbf{43.81} & \textbf{47.14} & \textbf{56.67} & \textbf{42.66} & \textbf{47.37} & \textbf{58.92} & \textbf{45.28} & \textbf{49.79} \\
\midrule
DePT         & 49.58 & 37.71 & 40.91 & 54.91 & 41.44 & 45.47 & 61.18 & 39.28 & 45.73 & 63.86 & 39.91 & 46.79 \\
\rowcolor{gray!20}
\textbf{+ SEPT}      & \textbf{50.28} & \textbf{41.82} & \textbf{44.15} & \textbf{55.29} & \textbf{41.45} & \textbf{45.98} & \textbf{61.73} & \textbf{41.03} & \textbf{47.56} & \textbf{64.57} & \textbf{41.63} & \textbf{49.06} \\
\bottomrule
\end{tabular}
}
\caption{Ablation of the number of shots per class. 
H denotes the harmonic mean between Base and New classes.
Reported values are averaged across 11 datasets.}
\label{tab:num_shots}
\end{table*}

\subsection{Details of Pengi.}

We utilize Pengi~\citep{pengi}, a multimodal model originally designed for audio-conditioned text generation. Structurally, it consists of three branches: an audio encoder, a text encoder, and a causal language model.
In line with prior work~\cite{palm}, we exclusively adopt the audio and text encoder branches to extract feature embeddings.
The audio branch is built upon the Hierarchical Token-Semantic Audio Transformer (HTSAT)~\citep{htsat}, adopting the architecture of CLAP~\cite{clap23}. The input processing pipeline involves specific preprocessing steps: raw audio waveforms are first resampled to 44.1 kHz and truncated or padded to a fixed duration of 7 seconds. These waveforms are then converted into log-Mel spectrograms before being mapped to the embedding space by the HTSAT encoder. In parallel, the text branch utilizes the CLIP~\citep{clip} transformer to generate corresponding linguistic embeddings.

\section{Additional Experiments}

\subsection{Ablation of the number of shots.}
To investigate the robustness of our method with respect to varying levels of supervision, we conduct an ablation study under different few-shot settings, where the number of training examples per seen class is set to 2, 4, 8, and 16. As shown in Table~\ref{tab:num_shots}, the proposed SEPT consistently improves generalization performance across all prompt tuning baselines and all shot settings.
Specifically, SEPT achieves notable gains in harmonic mean (H) under low-resource regimes. For instance, with only two shots per class, CoOp~\citep{coop} improves from 37.69\% to 41.83\%. This demonstrates that incorporating semantic neighbors provides a strong regularization effect even when labeled data is scarce. Similar trends are observed for 4-shot and 8-shot settings, where SEPT leads to stable improvements in both base and new class accuracy.
Even in the 16-shot setting, where models already achieve relatively strong performance, SEPT yields consistent gains. For example, DePT~\citep{dept} improves from 46.79\% to 49.06\% in H, and CoCoOp~\citep{cocoop} reaches 50.65\%, up from 46.26\%. These results highlight that SEPT enhances generalization not only in extremely low-shot regimes but also in moderately supervised scenarios, validating its effectiveness as a robust, plug-and-play augmentation for prompt tuning.

\begin{table}[t!]
\centering
\scalebox{0.9}{
\begin{tabular}{lccc}
\toprule
\multicolumn{1}{l}{\multirow{2}{*}{{Method}}} & \multicolumn{3}{c}{Avg over 11 datasets} \\
\cmidrule(l){2-4} 
 & {Base} & {New} & {H} \\ 
\midrule
Pengi      & 40.68 & 38.21 & 38.46 \\
CoOp       & 65.00 & 34.09 & 42.83 \\
\rowcolor{gray!20}
\textbf{+ SEPT$^{-}$}  & 63.37 & 36.58 & 45.09 \\
\rowcolor{gray!20}
\textbf{+ SEPT}       & 64.36 & 42.98 & 49.70 \\
\bottomrule
\end{tabular}
}
\caption{Ablation study on margin computation strategies. \textbf{SEPT$^{-}$} computes margins using a fixed prefix (``This is a sound of \{class\}''), while \textbf{SEPT} averages over a diverse set of $T=100$ hand-crafted prefixes. H denotes the harmonic mean between Base and New accuracy. Reported values are averaged across 11 datasets.
}
\label{tab:margin_computation_ablation}
\end{table}

\begin{table*}[ht!]
  \centering 
  \scalebox{0.9}{
    \begin{tabular}{lcccccccccccc}
      \toprule
     \multicolumn{1}{l}{\multirow{2}{*}{\makecell[c]{Method}}}  & \multicolumn{3}{c}{Avg over 11 datasets}  & \multicolumn{3}{c}{Beijing-Opera} & \multicolumn{3}{c}{NS-Instruments}  & \multicolumn{3}{c}{ESC50}  \\
    \cmidrule(lr){2-4}\cmidrule(lr){5-7}\cmidrule(lr){8-10}\cmidrule(lr){11-13}
    & Base & New & H & Base & New & H  & Base & New & H  & Base & New & H \\
    \midrule
    {CoOp}       & 65.00 & 34.09 & 42.83 & 97.27 & 61.38 & 74.87 & 41.11 & 23.88 & 30.08 & 61.97 & 11.83 & 19.46 \\
{KgCoOp}     & 37.99 & 37.42 & 36.39 & 67.26 & 61.40 & 62.96 & 39.80 & 39.27 & 39.11 & 14.10 & 10.33 & 11.76 \\
\midrule
{{KgCoOp\textsuperscript{\dag}} }     & \textbf{59.02} & 38.76 & 44.76 & \textbf{97.53} & \textbf{72.94} & \textbf{82.55} & 48.79 & \textbf{44.35} & \textbf{46.41} & \textbf{48.77} & 14.63 & 22.39 \\
\rowcolor{gray!20}
\textbf{+ SEPT}       & 58.34 & \textbf{40.52} & \textbf{46.31} & 97.23 & 71.22 & 81.41 & \textbf{49.66} & 42.04 & 45.48 & 46.47 & \textbf{15.77} & \textbf{23.39} \\

    \bottomrule
    \toprule
    \multicolumn{1}{l}{\multirow{2}{*}{\makecell[c]{Method}}}  & \multicolumn{3}{c}{ESC50-Actions}  & \multicolumn{3}{c}{UrbanSound8K} & \multicolumn{3}{c}{CREMA-D} & \multicolumn{3}{c}{RAVDESS}    \\
    \cmidrule(lr){2-4}\cmidrule(lr){5-7}\cmidrule(lr){8-10}\cmidrule(lr){11-13}
    & Base & New & H & Base & New & H  & Base & New & H  & Base & New & H  \\
    \midrule
    {CoOp}       & 85.33 & 49.33 & 61.97 & 61.86 & 31.22 & 40.94 & 52.50 & 9.01 & 15.29 & 51.01 & 27.75 & 35.88 \\
{KgCoOp}     & 36.50 & 30.67 & 32.88 & 26.14 & 26.15 & 25.40 & 42.70 & 55.43 & 42.48 & 27.53 & 27.46 & 27.42 \\
\midrule
{{KgCoOp\textsuperscript{\dag}} }     & \textbf{66.00} & 50.17 & 56.68 & \textbf{55.86} & 35.32 & 42.23 & \textbf{54.99} & 9.56 & 16.22 & \textbf{46.09} & 34.21 & \textbf{39.25} \\
\rowcolor{gray!20}
\textbf{+ SEPT}       & 65.33 & \textbf{53.17} & \textbf{58.37} & 55.50 & \textbf{36.48} & \textbf{43.10} & 51.12 & \textbf{19.34} & \textbf{26.15} & 43.56 & \textbf{34.65} & 38.21 \\

    \bottomrule
    \toprule
    \multicolumn{1}{l}{\multirow{2}{*}{\makecell[c]{Method}}}  & \multicolumn{3}{c}{SESA}  & \multicolumn{3}{c}{GT-Music-Genre} & \multicolumn{3}{c}{VocalSound}  & \multicolumn{3}{c}{TUT2017}  \\
    \cmidrule(lr){2-4}\cmidrule(lr){5-7}\cmidrule(lr){8-10}\cmidrule(lr){11-13}
    & Base & New & H & Base & New & H  & Base & New & H  & Base & New & H   \\
    \midrule
    {CoOp}       & 85.78 & 74.77 & 79.33 & 57.76 & 20.54 & 29.76 & 75.26 & 45.01 & 56.00 & 45.11 & 20.25 & 27.53 \\
{KgCoOp}     & 68.14 & 79.28 & 72.37 & 27.39 & 15.82 & 19.77 & 52.50 & 49.91 & 51.10 & 15.83 & 15.90 & 15.10 \\
\midrule
{{KgCoOp\textsuperscript{\dag}} }     & 78.43 & 77.48 & 77.94 & \textbf{50.83} & 19.53 & 28.18 & 68.30 & 52.06 & 59.08 & 33.62 & 16.11 & 21.50 \\
\rowcolor{gray!20}
\textbf{+ SEPT}       & \textbf{79.41} & \textbf{78.38} & \textbf{78.74} & 49.50 & \textbf{21.89} & \textbf{30.04} & \textbf{69.79} & \textbf{55.73} & \textbf{61.96} & \textbf{34.20} & \textbf{17.06} & \textbf{22.50} \\

    \bottomrule
    \end{tabular}
}
\caption{Comparison of base-to-new generalization across different methods with and without SEPT. 
KgCoOp uses a single prompt template (''This is a sound of'') as in the original paper, 
whereas KgCoOp\textsuperscript{\dag} averages predictions over 100 hand-crafted prompt templates 
to enhance robustness and generality. All methods are evaluated using 16-shot learning.
}
\label{tab:b2n_kgcoop}
\end{table*}

\subsection{Ablation of Margin Computation Strategy.}
Table~\ref{tab:margin_computation_ablation} presents an ablation study comparing two margin computation strategies in our semantic expansion loss. In SEPT$^{-}$, we compute all semantic margins using a single fixed prefix template: {``This is a sound of \{class\}''}. This approach already brings noticeable improvements over the CoOp baseline, demonstrating the effectiveness of semantic alignment through even a single prompt template.
However, relying solely on one fixed prefix may introduce bias or noise, especially if the template happens to be semantically ambiguous or suboptimal for certain classes. To mitigate this, we propose averaging distances across a diverse pool of $T=100$ hand-crafted prefixes. This strategy smooths out template-specific noise and produces a more robust estimation of the semantic margin.
As shown in the table, SEPT achieves a substantial performance boost, improving the harmonic mean from 45.09\% to 49.70\%. This result validates our hypothesis that averaging over multiple semantic views leads to a more stable and generalizable prompt embedding space.

\subsection{Analysis of KgCoOp.}
\label{appendix:kgcoop_analysis}
Contrary to the performance improvement trend of KgCoOp~\citep{kgcoop} in vision-language models (VLMs), our experiments in the main paper show that it performs significantly worse than CoOp. 
To further investigate this discrepancy, we conduct a detailed analysis using Table~\ref{tab:b2n_kgcoop}.

Following the original implementation, we adopt a single hand-crafted prompt template—{``This is a sound of {class}''}—to compute the zero-shot prompt embedding, which serves as the knowledge-guided regularizer for the learned prompt.
However, across all datasets, KgCoOp underperforms not only relative to other prompt tuning methods such as CoOp, but in some cases even falls below the zero-shot baseline.
This performance drop can be attributed to KgCoOp’s heavy reliance on the quality of the zero-shot prompt embedding. Since the method is designed to preserve knowledge from the fixed prompt, it implicitly assumes that the hand-crafted prompt yields a robust and semantically rich representation.
In our setting, however, the fixed prompt often fails to capture the diverse acoustic semantics of different audio classes, resulting in poor zero-shot performance.
While this assumption may hold in VLMs—where strong pretrained models like CLIP and rich textual supervision lead to generally reliable zero-shot regularizer—it becomes less valid in the ALM setting. Compared to vision-language tasks, audio-language models often operate in a narrower semantic space with fewer class labels, and many audio classes are not easily described by generic prompts such as {``This is a sound of''}. As a result, a poorly aligned base class regularizer can negatively affect transferability and generalization in ALMs.

To address this issue, we construct an enhanced version, KgCoOp\textsuperscript{\dag}, which averages the zero-shot embeddings obtained from 100 diverse hand-crafted prompt templates. This ensembling strategy yields a more robust and semantically stable regularizer by mitigating the bias introduced by any single prompt.
As a result, KgCoOp\textsuperscript{\dag} achieves substantial performance improvements across most datasets, confirming that the diversity and quality of prompts used for knowledge anchoring are critical to the success of knowledge-guided prompt tuning.
Furthermore, we apply our proposed SEPT framework on top of KgCoOp\textsuperscript{\dag}, further enhancing the generalization ability of the model. SEPT regularizes the prompt embedding space by leveraging semantic neighbors, which complements the stability provided by prompt ensembling.

\begin{figure}[t!]
  \centering
    \includegraphics[width=\linewidth]{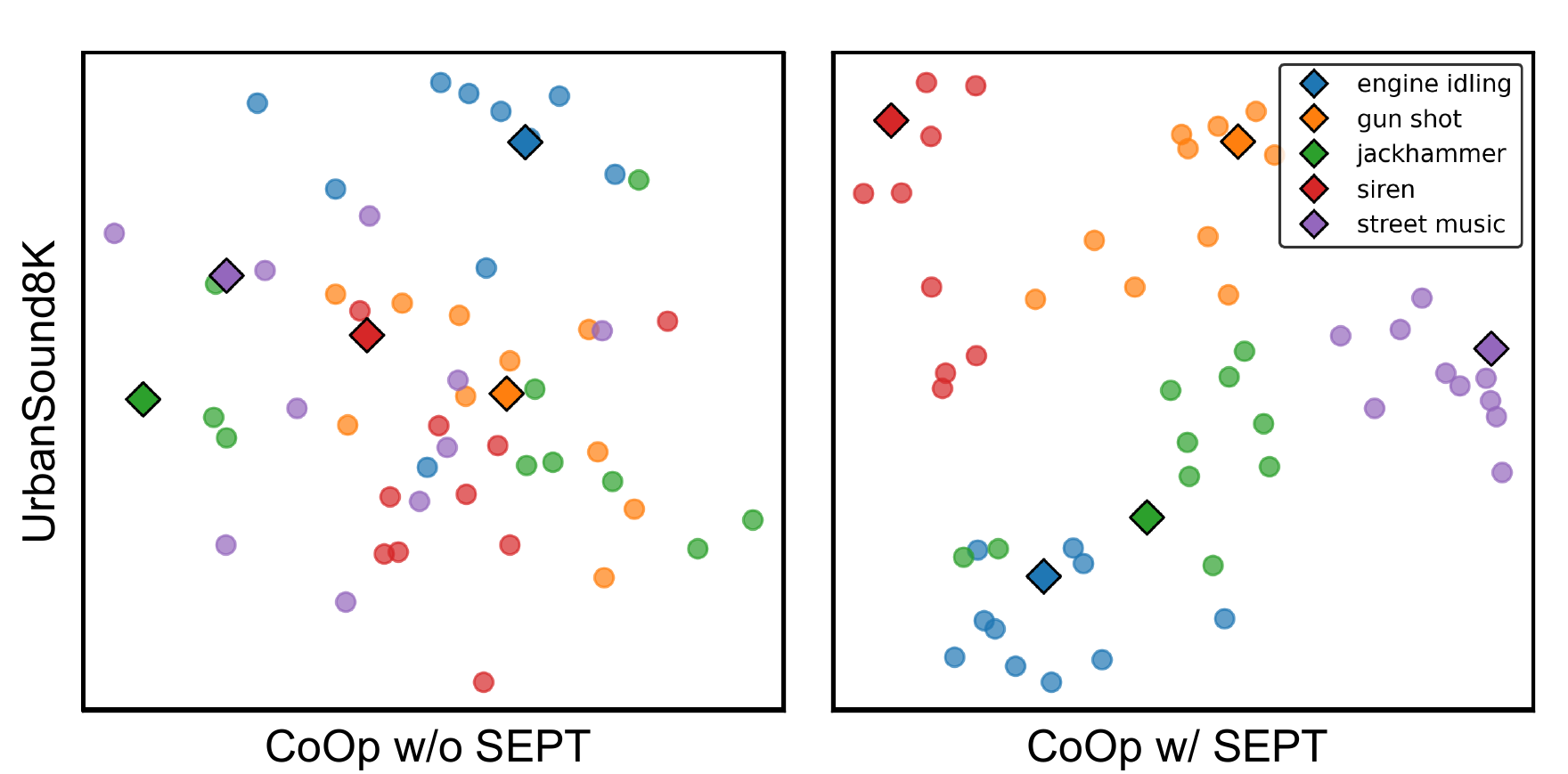}
    \includegraphics[width=\linewidth]{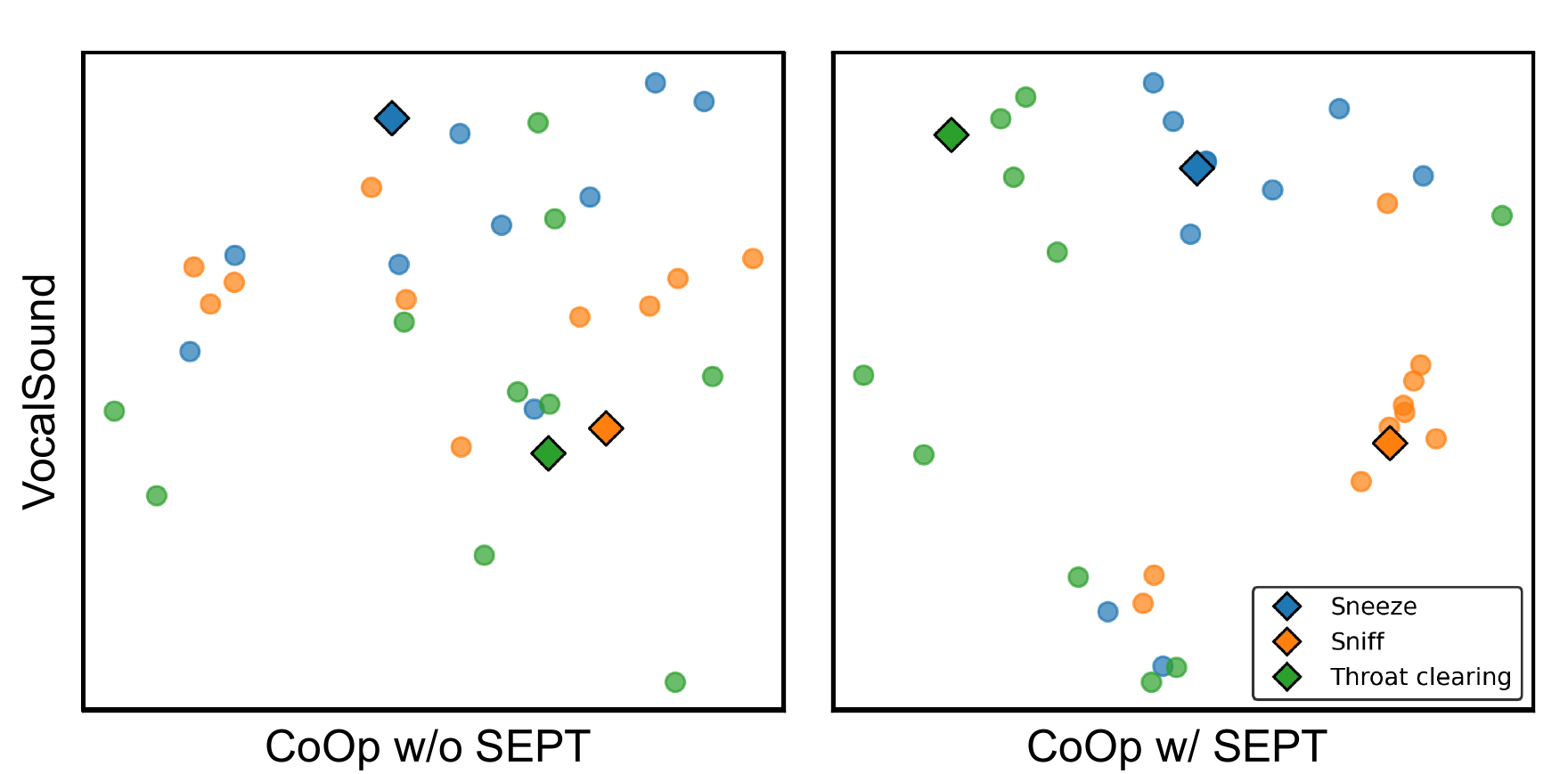}
  \caption{t-SNE visualization of prompt embeddings of new classes and their semantic neighbors from UrbanSound8K dataset (top), VocalSound dataset (bottom).
  $\diamond$ denotes the prompt embedding of a new class, while $\circ$ represents its corresponding semantic neighbors.
  }
  \label{fig:additional_tsne}
\end{figure}

\subsection{Additional Qualitative Results.}
\label{appndix:additional_qualitative}
We provide additional t-SNE visualizations of prompt embeddings for new classes and their semantic neighbors from the UrbanSound8K~\citep{urbansound} and VocalSound~\citep{vocalsound} datasets in Fig.~\ref{fig:additional_tsne}.
Importantly, the new classes and their corresponding neighbors were not used during training to enforce any semantic structure.
Nonetheless, consistent with the observations in the main paper, SEPT produces a well-structured embedding space in which new classes are closely clustered with their semantic neighbors. These results further demonstrate the strong generalization capability of our method across diverse audio classification tasks.

\begin{figure}
    \centering
    \includegraphics[width=\linewidth]{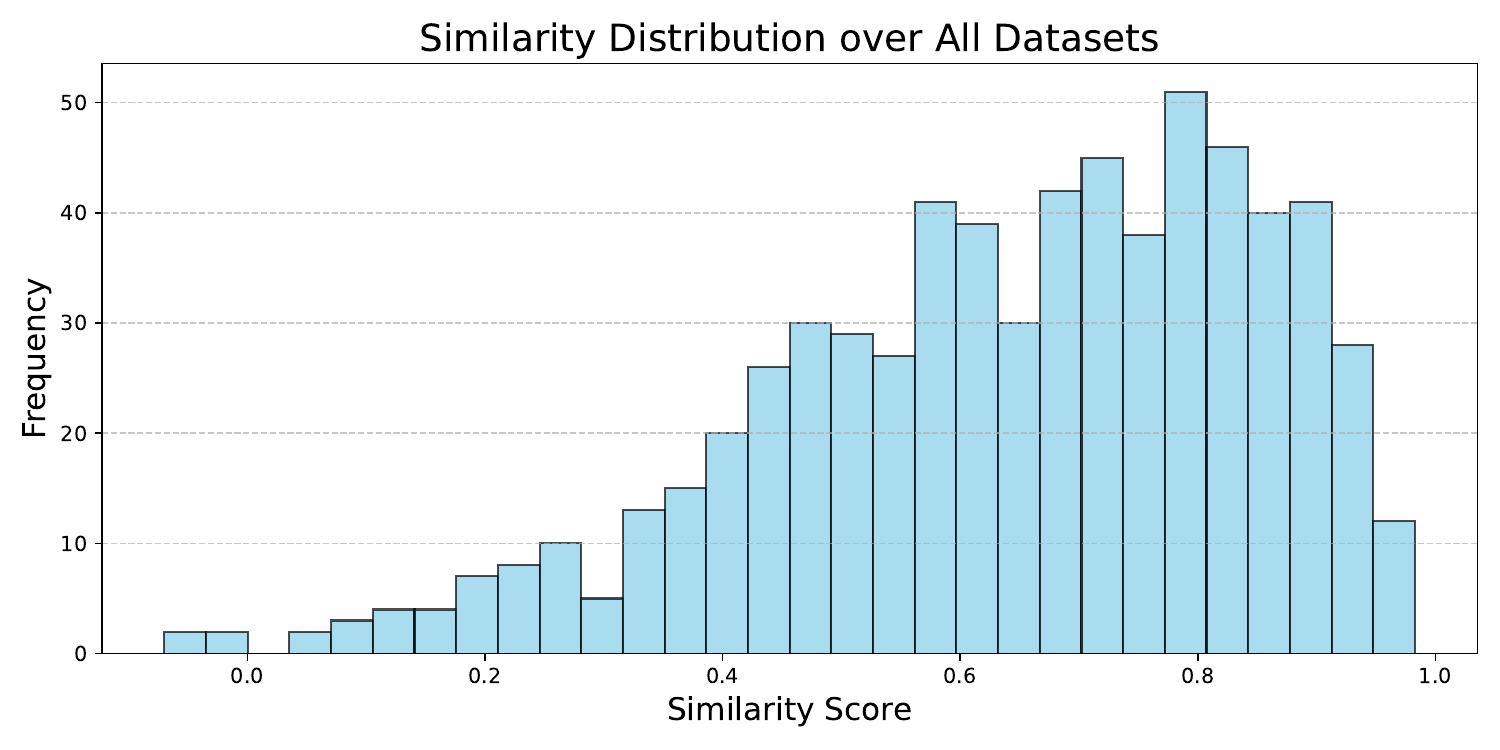}
    \caption{Distribution of similarity scores between classes and semantic neighbors across all datasets. The majority of similarity scores fall between 0.6 and 0.9, indicating a tendency toward moderately high semantic similarity among features.}
    \label{fig:similarity_distribution}
\end{figure}

\subsection{Quality of Semantic Neighbors.}
\label{sec:neighbor_quality}
Our framework relies on semantic neighbors generated by large language models (LLMs), such as ChatGPT-4o~\citep{gpt4o}, to enhance the semantic structure of the prompt embedding space.
Ideally, these semantic neighbors should remain semantically close to the original class while being sampled with slight variations to ensure diversity, which is crucial for achieving effective semantic expansion.
To assess the quality of the generated neighbors, we evaluated both their similarity to the original class and their internal diversity using a text encoder with a hand-crafted prompt (''This is a sound of'').

\begin{table}[t]
\centering
\scalebox{0.8}{
\begin{tabular}{l c}
\toprule
\textbf{Dataset} & \textbf{Mean Diversity} \\ \midrule
Beijing-Opera    & 0.4482 \\
NS-Instruments   & 0.4964 \\
ESC50            & 0.4296 \\
ESC50-Actions    & 0.4141 \\
UrbanSound8K     & 0.5657 \\
CREMA-D          & 0.4480 \\
RAVDESS          & 0.4618 \\
SESA             & 0.4382 \\
GT-Music-Genre   & 0.3491 \\
VocalSound       & 0.5217 \\
TUT2017          & 0.3294 \\ \midrule
Average          & 0.4457 \\
\bottomrule
\end{tabular}
}
\caption{Mean diversity for each dataset.}
\label{tab:diversity}
\end{table}

In Fig.~\ref{fig:similarity_distribution}, we visualize the distribution of cosine similarity scores between each class and its semantic neighbors
across all datasets.
The majority of semantic neighbors exhibit high similarity scores, indicating that most are well-aligned and semantically relevant. However, a small number show notably low similarity, suggesting the presence of noisy or overly generic neighbors.
Furthermore, we present the diversity score of each dataset to quantify how diversely the semantic neighbors are sampled within the original classes in Table~\ref{tab:diversity}. The diversity of class $c_i$ is formulated as:
\begin{equation}
\text{Diversity}_i 
= 1 - \frac{1}{{}_NC_2} 
\sum_{n<m} 
\operatorname{sim}(\mathbf{p}_{i}^{n}, \mathbf{p}_{i}^{m}),
\end{equation}
where $\operatorname{sim}(\cdot , \cdot)$ is the cosine similarity.
For each dataset, we computed the diversity score for each class and then averaged them across all classes.
On average, a diversity score of approximately 0.45 was achieved, indicating a moderate level of semantic variability among neighbors within each class.
Overall, the results indicate that the semantic neighbors maintain strong relevance to their original classes while providing enough variation to enhance prompt tuning performance.

\subsection{Comparison with Alternative Topology-Preserving Mechanisms}
\label{sec:alternative_mechanisms}

To further validate the necessity of our proposed localized semantic expansion, we investigate whether alternative topology-preserving mechanisms or global regularization strategies could achieve similar improvements in base-to-new generalization. Specifically, we implement and compare two additional baselines. First, inspired by CLIP-Adapter, we introduce Text-Adapter, a learnable linear layer (adapter) on top of the frozen text embeddings. This topology-preserving mechanism adapts the representation without explicitly reshaping local neighborhood relations via loss constraints. Second, we explore Global Regularization, where instead of using targeted local semantic neighbors, we regularize the text embeddings against large, global concept pools via an L2 loss. For this approach, we evaluate two variants, each utilizing a pool of 200 concepts: (1) Global Reg. (FSD50K-200), which consists of classes sampled from the FSD50K dataset (a subset of PENGI's pretraining data), and (2) Global Reg. (LLM-random-200), which consists of random audio-related concepts generated by an LLM.

\begin{table}[t!]
\centering
\scalebox{0.8}{
\begin{tabular}{lccc}
\toprule
\textbf{Method} & \textbf{Base} & \textbf{New} & \textbf{H} \\
\midrule
CoOp & \textbf{65.00} & 34.09 & 42.83 \\
Text-Adapter & 60.85 & 41.30 & 46.67 \\
Global Reg. (FSD50K-200) & 64.74 & 39.21 & 46.81 \\
Global Reg. (LLM-random-200) & 63.32 & 41.66 & 47.50 \\
\midrule
\textbf{SEPT (Ours)} & 64.36 & \textbf{42.98} & \textbf{49.70} \\
\bottomrule
\end{tabular}
}
\caption{Comparison with alternative topology-preserving and global regularization mechanisms. Performance is averaged across 11 datasets.}
\label{tab:alternative_mechanisms}
\end{table}

As shown in Table~\ref{tab:alternative_mechanisms}, both the topology-preserving Text-Adapter and the global regularization strategies improve new-class generalization compared to the standard CoOp baseline. This indicates that preventing the structural collapse of the pre-trained embedding space---either via architectural constraints or global anchors---is indeed beneficial.
However, our proposed SEPT still consistently outperforms these alternatives, achieving the highest harmonic mean (49.70\%). This demonstrates that the gains from alternative topology-preserving methods are complementary to our approach. The targeted, task-relevant inductive bias provided by localized semantic neighbor expansion in SEPT proves to be more effective for base-to-new generalization than relying solely on linear adapters or broad global concept pools.

\subsection{Applicability on VLMs.}

To further support the plug-and-play property of SEPT beyond ALMs, we additionally evaluate SEPT on a VLM prompt tuning setup. Specifically, we apply SEPT on CoOp, using CLIP \citep{clip} as a backbone, and evaluate the base-to-new generalization on Caltech101 \citep{caltech101}. As shown in Table~\ref{tab:sept_vlm_caltech101}, SEPT improves performance on both base and novel classes, indicating that it transfers effectively to VLM prompt tuning.
These results validate the modality-agnostic nature of SEPT, confirming its potential as a universal regularizer that enhances prompt tuning performance across both audio and vision domains.
However, we emphasize that the impact and practicality of SEPT are particularly pronounced in the audio domain. Unlike vision benchmarks, which often contain dense category sets (e.g., ImageNet), audio datasets frequently suffer from class sparsity. In such sparse semantic spaces, the prompt embedding structure is more prone to degradation, making the semantic expansion provided by SEPT even more critical for preserving generalization capabilities. Furthermore, since the computational cost of neighbor-based regularization scales with the number of categories, SEPT remains highly efficient in these sparse audio settings, offering a favorable balance between structural robustness and training overhead compared to dense vision tasks.

\begin{table}[t]
\centering
\scalebox{0.85}{
\begin{tabular}{lccc}
\toprule
Method & Base & New  & H \\
\midrule
CoOp            & 98.03 & 90.10 & 93.90 \\
\rowcolor{gray!20}
+ \textbf{SEPT}     & \textbf{98.23} & \textbf{93.40} & \textbf{95.76} \\
\bottomrule
\end{tabular}
}
\caption{Results on Caltech101 under the base-to-new generalization protocol with CLIP/CoOp. SEPT consistently improves base and novel accuracies and increases the harmonic mean.}
\label{tab:sept_vlm_caltech101}
\end{table}

\section{Additional Analysis}

\subsection{Analysis of embedding space structure collapse in ALMs and VLMs}
To clarify why SEPT is particularly beneficial in the audio domain, we analyze how prompt tuning changes the global structure of the base-class text embedding space in both ALMs and VLMs. Starting from the zeroshot text prototypes and then applying CoOp, we measure two complementary indicators of global geometry: (i) the mean pairwise cosine similarity between class embeddings, which reflects how strongly class prototypes become globally clustered, and (ii) the normalized matrix entropy of the singular-value spectrum, which captures how evenly the embedding energy is distributed across dimensions. A lower matrix entropy indicates a more anisotropic and lower-dimensional geometry, i.e., stronger collapse.

Let $Z \in \mathbb{R}^{K \times d}$ denote the matrix of base-class text embeddings, where $K$ is the number of base classes and $d$ is the embedding dimension. Given the singular values $\{s_i\}_{i=1}^{r}$ of $Z$, we define $p_i = s_i / \sum_{j=1}^{r} s_j$ and compute the normalized matrix entropy as
\[
H_{\mathrm{norm}}(Z) = \frac{-\sum_{i=1}^{r} p_i \log p_i}{\log r},
\]
where $r$ is the rank of $Z$.

\begin{table}[t]
\centering
\small
\scalebox{0.75}{
\begin{tabular}{llcccccc}
\toprule
\multirow{2}{*}{Domain} & \multirow{2}{*}{Dataset} & \multicolumn{2}{c}{Cos. Sim.} & \multicolumn{2}{c}{Mat. Ent.} & \multirow{2}{*}{\# Base} \\
\cmidrule(lr){3-4} \cmidrule(lr){5-6}
& & Zeroshot & CoOp & Zeroshot & CoOp & \\
\midrule
Audio & CREMA-D    & 0.583 & 0.921 & 0.911 & 0.664 & 3 \\
Audio & ESC50      & 0.363 & 0.505 & 0.942 & 0.882 & 25 \\
Image & Caltech101 & 0.696 & 0.404 & 0.907 & 0.953 & 50 \\
Image & ImageNet   & 0.544 & 0.538 & 0.892 & 0.883 & 500 \\
\bottomrule
\end{tabular}
}
\caption{Comparison of global geometry changes in the base-class text embedding space after prompt tuning. In ALMs, CoOp increases global clustering and decreases effective dimensionality, whereas VLMs show much weaker or no collapse-like behavior.}
\label{tab:alm_vlm_collapse}
\end{table}

Table~\ref{tab:alm_vlm_collapse} summarizes the results. In ALMs, prompt tuning consistently increases the mean pairwise cosine similarity while decreasing matrix entropy, indicating that class prototypes become more globally clustered and concentrated in a lower-dimensional subspace. This trend is especially pronounced in semantically sparse audio benchmarks such as CREMA-D and ESC50. In contrast, VLMs do not exhibit the same collapse-like behavior: the cosine similarity remains stable or even decreases, and the matrix entropy is largely preserved. These observations suggest that prompt tuning distorts the pretrained text-space geometry much more severely in ALMs than in VLMs. We attribute this difference to the fact that audio benchmarks typically contain fewer and sparser semantic categories, making the prompt space more vulnerable to structural degradation. This also explains why preserving local semantic structure is particularly important in ALM few-shot adaptation, while SEPT remains applicable to VLMs as a general regularization strategy.

\subsection{Neighbor Quality.}

To assess the reliability of the generated semantic neighbors, we quantify their alignment with class concepts by computing the average cosine similarity across 11 datasets. The results—0.64 for positive pairs versus 0.34 for negative pairs—demonstrate that the generated neighbors achieve a balance between semantic coherence and diversity. Qualitative inspection confirms that these neighbors capture diverse facets of a class (e.g., neighbors for footsteps include specific variations like running steps or gravel crunch).

\begin{table}[t]
\centering
\scalebox{0.9}{
\begin{tabular}{lccc}
\toprule
\multicolumn{1}{l}{\multirow{2}{*}{{Method}}} & \multicolumn{3}{c}{Avg over 11 datasets} \\
\cmidrule(l){2-4} 
 & {Base} & {New} & {H} \\ 
\midrule
CoOp                   & 65.00 & 34.09 & 42.83 \\
\rowcolor{gray!20}
+ \textbf{SEPT}        & 64.36 & 42.98 & {49.70} \\
\rowcolor{gray!20}
+ \textbf{SEPT\textsuperscript{\ddag}}        & 64.37 & 42.93 & {49.64} \\

\bottomrule
\end{tabular}
}
\caption{Impact of new class name overlap in generated semantic neighbors. Filtering these neighbors (SEPT\textsuperscript{\ddag}) yields nearly identical performance over 11 datasets, indicating minimal effect from the overlap.}
\label{tab:leakge}
\end{table}

As shown in Fig.~\ref{fig:similarity_distribution} and Table~\ref{tab:diversity}, certain semantic neighbors exhibit low similarity to their target classes or lack sufficient diversity, which may introduce noise or redundancy into the embedding space.
Crucially, SEPT exhibits robustness to potential generation noise because neighbors are employed exclusively for text-space regularization, distinct from the cross-entropy supervision used for base class classification. Consequently, noisy neighbors do not directly corrupt the primary decision boundaries. The core safeguard lies in our margin constraint, which is precomputed based on the text-space geometry of the pretrained ALM—a space already aligned with audio semantics via contrastive pretraining. In this framework, an acoustically distant positive neighbor is assigned a large margin, effectively preventing the loss term from activating unless the embedding violates this geometric prior. This mechanism acts as a gate for gradient updates, naturally capping the influence of outliers and mitigating the impact of neighbor noise.

We further analyze whether the generated neighbors contain terms that overlap with the held-out new class names, to address the potential concern that such overlap could inadvertently introduce new class information.
Across the 11 datasets, only 1.7\% of the neighbors generated for base classes overlap with any new class name, indicating that such cases are rare.
More importantly, this overlap does not imply using any new class audio samples: SEPT never accesses new class audio samples during training, and the neighbors are not used in the cross-entropy objective for base class classification.
Instead, neighbors are used solely in the text-space regularization term, which shapes the prompt embedding geometry rather than directly supervising class decision boundaries.
To empirically assess the impact of this overlap, we define \textbf{SEPT\textsuperscript{\ddag}}, which filters out all neighbors that overlap with any new class name before training.
As shown in Table~\ref{tab:leakge}, SEPT\textsuperscript{\ddag} achieves performance nearly identical to SEPT over 11 datasets, suggesting that the observed gains are not driven by the small fraction of overlapping neighbors.

\subsection{Performance Variations per Dataset.}
While SEPT consistently improves generalization across the majority of benchmarks, we observe marginal gains or slight performance drops in specific dataset--method combinations. We attribute these variations to two primary factors:

\textbf{1. Semantic Ambiguity:}
In domains with high abstraction, such as music genres or emotional states, semantic boundaries are inherently ambiguous. In such cases, generated neighbors may overlap between classes or lack distinctiveness (e.g., exhibiting small similarity gaps between positive and negative neighbors). This can introduce noise into the embedding space rather than structural clarity, potentially hindering the precise optimization of base classes.

\textbf{2. Compound Regularization Effects:} SEPT functions as an explicit regularizer to structure the embedding space. When integrated with baseline methods that already impose strong constraints (e.g., via knowledge distillation or channel bias correction), the addition of SEPT may result in an over-regularized objective. This can excessively constrain the optimization landscape, limiting the model's ability to fully fit the base classes. Nevertheless, even in these instances, SEPT typically preserves or improves the harmonic mean, demonstrating that the trade-off favors enhanced generalization to unseen classes.

\section{AI Assistant Usage Statement}

We utilize Large Language Models (LLMs) for two distinct purposes in this study:

\paragraph{Model-based Experimentation.}
We employ ChatGPT-4o as the primary engine to generate semantic neighbors for the proposed SEPT framework. Additionally, to verify the robustness of our method across different architectures (as detailed in Sec.~\ref{sec:llm_dependency}), we utilize Gemini 2.5 Pro and Qwen3 Max to generate neighbor candidates.

\paragraph{Writing Assistance.}
We use Gemini 2.5 Pro exclusively for grammatical error correction and stylistic polishing of the manuscript. We confirm that no AI tools were used to generate scientific claims, experimental results, or the core intellectual contributions of this paper.

\clearpage

\end{document}